\documentclass{emulateapj} 
\usepackage{psfig}
\usepackage{txfonts}
\usepackage{bm}
\usepackage[usenames,dvipsnames]{color}
\usepackage{graphicx}

\newcommand{\unitstyle}[1]{\ensuremath{\mathrm{#1}}}

\newcommand{\second}{\unitstyle{s}}

\newcommand{\Om}{\ensuremath{\Omega}}  
\newcommand{\Omc}{\ensuremath{\Om_{\mathrm{crit}}}} 
\newcommand{\tkh}{\ensuremath{\tau_{\mathrm{KH}}}} 

\newcommand{\Ledd}{\ensuremath{L_{\mathrm{Edd}}}} 
\newcommand{\Mdot}{\ensuremath{\dot{M}}} 

\newcommand{\Msun}{\ensuremath{\unitstyle{M}_\odot}}

\newcommand{\kms}{\ensuremath{\mathrm{km}\,\second^{-1}}}
\newcommand{\vcrit}{{\varv_{\mathrm{crit}}}}

\newcommand{\vsurf}{{\varv_{\mathrm{surf}}}}
\newcommand{\vsurfi}{{\varv_{\mathrm{surf, ini}}}}
\newcommand{\alphasc}{\ensuremath{\alpha_{\mathrm{sc}}}} 

\newcommand{\nablaad}{\ensuremath{\nabla_{\!\mathrm{ad}}}}	
\newcommand{\nablaT}{\ensuremath{\nabla_{\!T}}}	
\newcommand{\supernab}{\ensuremath{\delta_\nabla}}  

\begin{document}
\title{The fate of fallback matter around newly born compact objects}

\author{Rosalba Perna\altaffilmark{1,2}, Paul Duffell\altaffilmark{3}, 
Matteo Cantiello\altaffilmark{4}, Andrew I. MacFadyen\altaffilmark{3}}

\affil{$^1$JILA \& Department of Astrophysical and Planetary Sciences,
  University of Colorado, 440 UCB, Boulder, CO80309-0440, USA;
  rosalba@jilau1.colorado.edu} 

\affil{$^2$ Department of Physics and Astronomy, Stony Brook University, Stony Brook, NY 11794-3800, USA}

\affil{$^3$ Center for Cosmology and Particle Physics, Department of
  Physics, New York University, 4 Washington Place, New York, NY
  10003, USA}

\affil{$^4$ Kavli Institute for Theoretical Physics, University of
  California, Santa Barbara, Kohn Hall, CA 93106, USA}

\begin{abstract}

The presence of fallback disks around young neutron stars has been
invoked over the years to explain a large variety of phenomena.  Here
we perform a numerical investigation of the formation of such disks
during a supernova explosion, considering both neutron star (NS) and
black hole (BH) remnants. Using the public code MESA, we compute the
angular momentum distribution of the pre-supernova material, for stars
with initial masses $M$ in the range $13-40~M_\odot$, initial surface rotational
velocities $\vsurf$ between 25\% and 75\% of the critical velocity, and for
metallicities $Z$ of 1\%, 10\% and 100\% of the solar value.  These
pre SN models are exploded with energies $E$ varying between $10^{50}-3\times
10^{52}$~ergs, and the amount of  fallback material is computed.  We find that,
if magnetic torques play an important role in angular momentum
transport, then fallback disks around NSs, even for low-metallicity
main sequence stars, are {\em not} an outcome of SN
explosions. Formation of such disks around young NSs can only happen
under the condition of negligible magnetic torques and a fine-tuned
explosion energy.  For those stars which leave behind BH remnants,
disk formation is ubiquitous if magnetic fields do not play a strong
role; however, unlike the NS case, even with strong magnetic coupling
in the interior, a disk can form in a large region of the {$Z,M,\vsurf,E$}
parameter space. Together with the compact, hyperaccreting fallback
disks widely discussed in the literature, we identify regions in the
above parameter space which lead to extended, long-lived disks around
BHs.  We find that the physical conditions in these disks may be
conducive to planet formation, hence leading to the possible existence
of planets orbiting black holes.

\end{abstract}

\keywords{stars: evolution --- stars: neutron --- supernovae: general --- accretion, accretion disks}

\section{Introduction}

The presence of fallback associated with supernova (SN) explosions
was recognized since early SN studies (Colgate 1971; Chevalier 1989).
It occurs when part of the ejecta fails to achieve terminal
  escape velocity and accretes onto the central compact object.  If
some of this material has angular momentum larger than the
keplerian value at the last stable orbit (or at the surface of the compact
object, if a neutron star), then the formation of a disk is expected.

Fallback disks around newly born compact objects have been invoked to
explain a large variety of phenomena.  Hyperaccreting disks formed
from the fallback of massive, rapidly rotating stars, whose iron cores
collapse into black holes, are believed to provide the energy source
that powers the class of long-duration Gamma-Ray Bursts (GRBs)
(Woosley 1993, MacFadyen \% Woosley 1999). MacFadyen \& Woosley (1999)
studied the formation and evolution of such a disk numerically,
following the explosion of a 35 $M_\odot$ main sequence star with
angular momentum in the $3\times 10^{16}-2\times 10^{17}$ cm$^2$
s$^{-1}$ range.  The detailed angular momentum structure of the
fallback material is believed to play a role in the early X-ray
emission following the prompt GRB (e.g. Kumar et al. 2008; Cannizzo
\& Gehrels 2009; Perna \& MacFadyen 2010; Geng et al. 2013).

\begin{figure*}
\begin{minipage}[t]{0.5\textwidth}
\centering
\includegraphics[width=8.2cm]{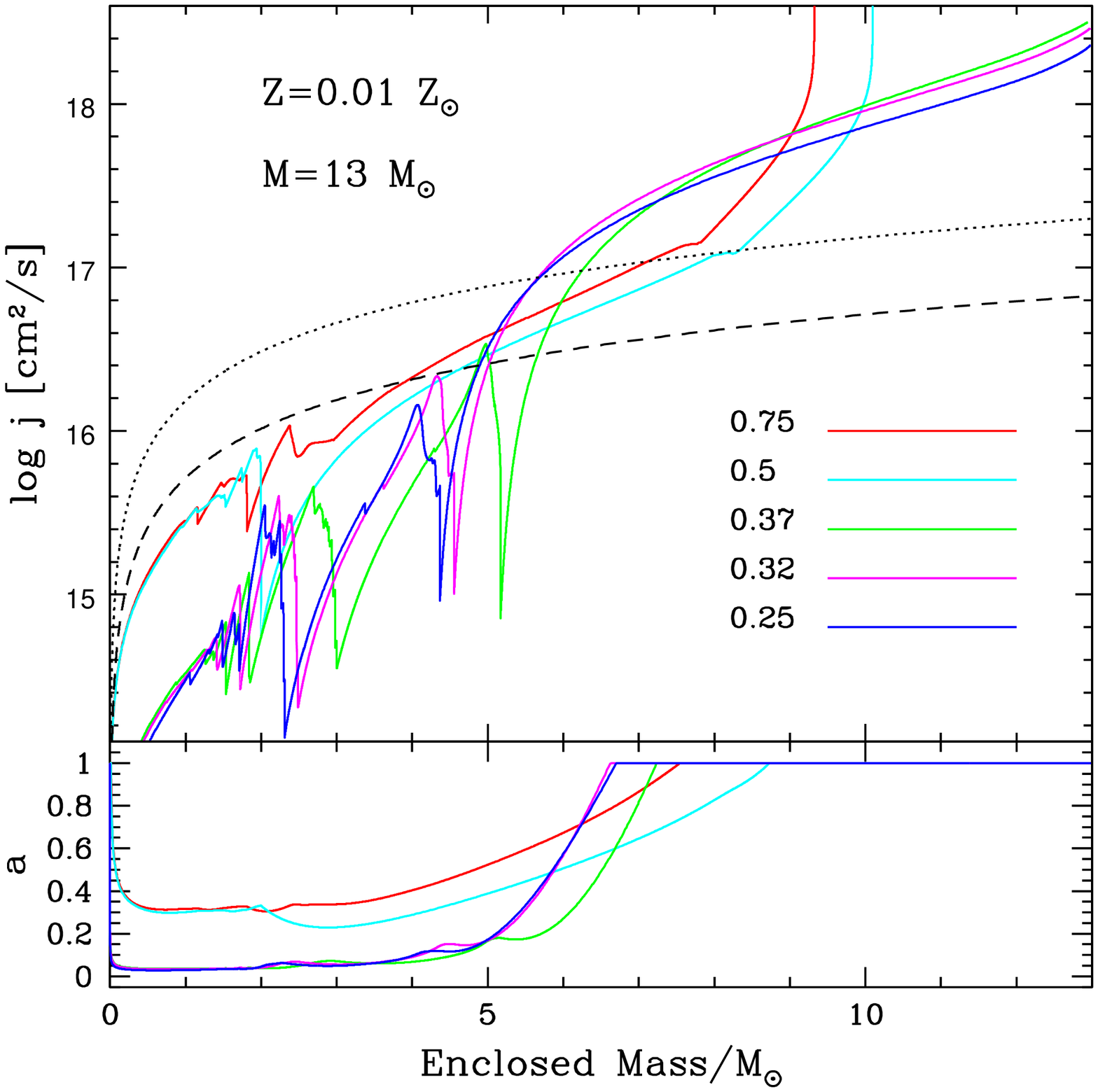}
\end{minipage}
\begin{minipage}[t]{0.5\textwidth}
\centering
\includegraphics[width=8.2cm]{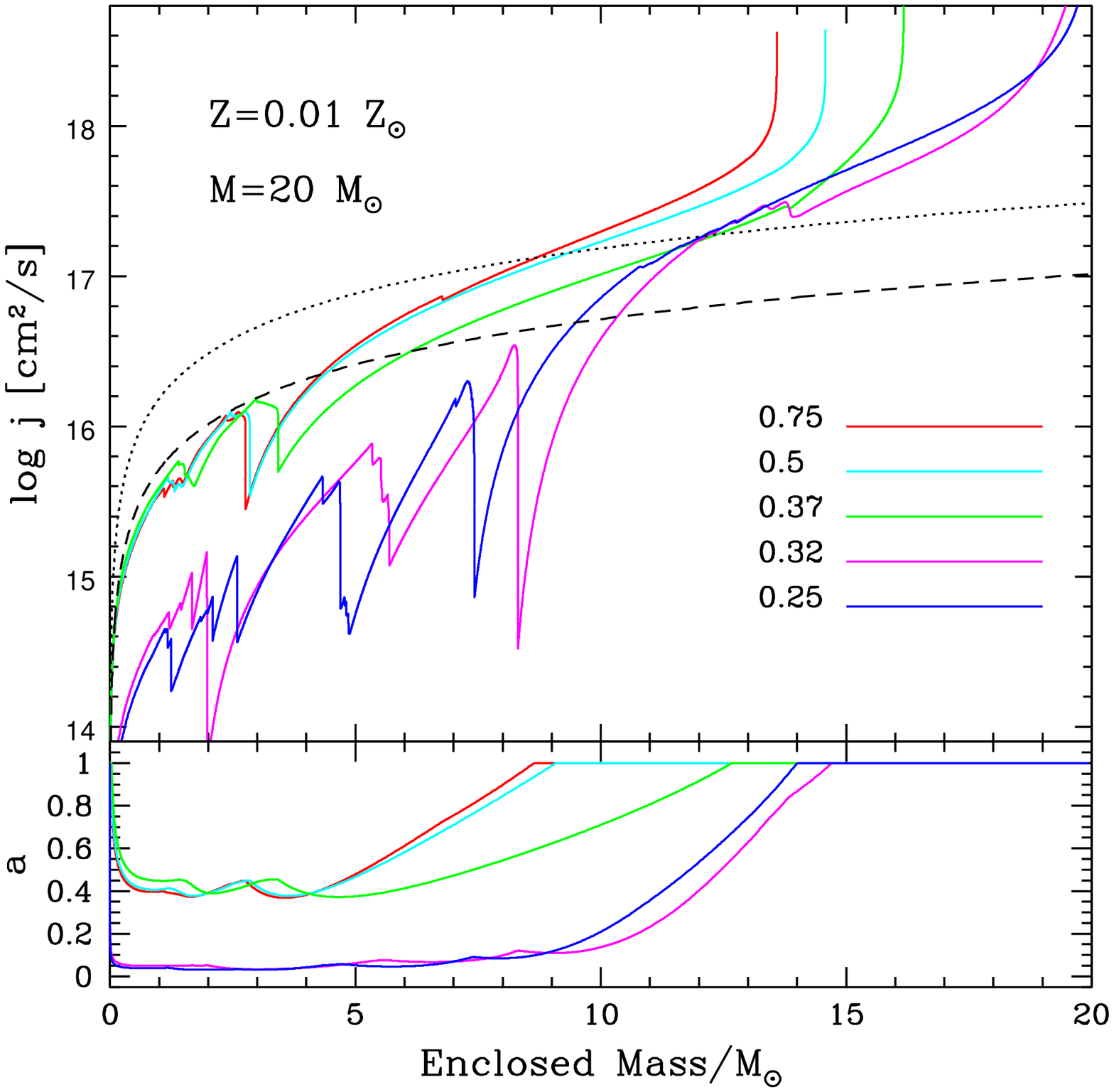}
\end{minipage}
\begin{minipage}[t]{0.5\textwidth}
\centering
\includegraphics[width=8.2cm]{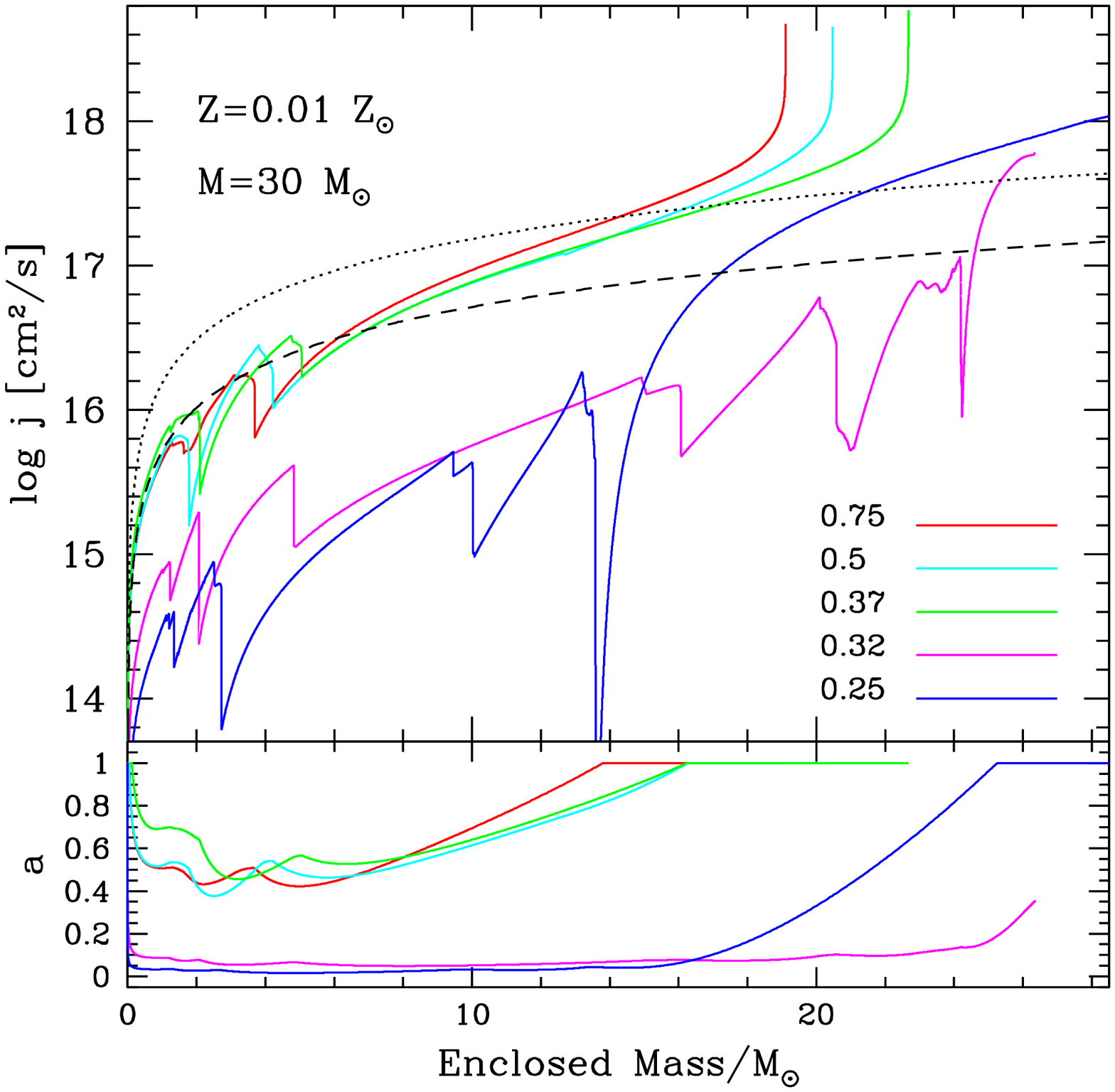}
\end{minipage}
\begin{minipage}[t]{0.5\textwidth}
\centering
\includegraphics[width=8.2cm]{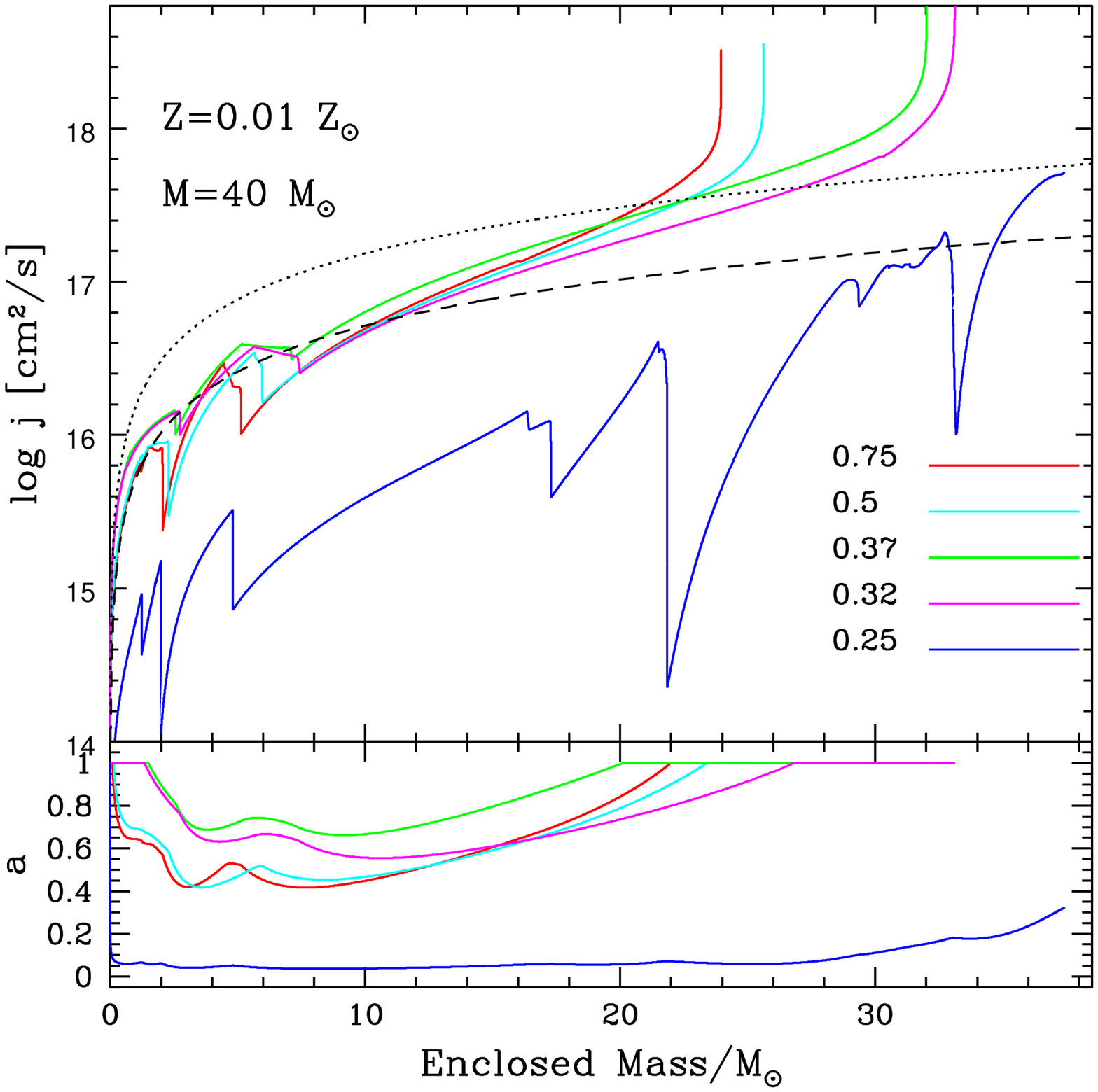}
\end{minipage}
\caption{The distribution of specific angular momentum in the pre-SN
  star for main sequence stars of 1\% solar metallicity and a range of
  initial surface rotational velocities, expressed in units of the critical
  velocity. Dashed and dotted lines in the upper region of each panel
  represent, respectively, the specific angular momentum of the last
  stable orbit around a Schwarzschild black hole and a maximally
  rotating Kerr black hole (of mass equal to the enclosed mass). The
  spin parameter of the enclosed material is shown, for each pre-SN
  model, in the bottom region of each panel.}
\label{fig:preSN1}
\end{figure*}

In the case of neutron stars (NSs), there is a much larger variety of
phenomena that has been potentially ascribed to the presence of a
fallback disk. Michel \& Dessler (1981) and Michel (1988) originally
emphasized that the presence of such disks can play a role in the
evolution of young NSs. Lin, Woosley \& Bodenheimer (1991)
noticed how such disks can provide a clue regarding the formation of
planets around young neutron stars, while Chatterjee, Hernquist \&
Narayan (2000), and Alpar (2001) showed how fallback disks around
NSs of ages $\sim 10^4-10^5$ yr can produce, by accretion
onto the NS surface, X-ray luminosities as high as $\sim
10^{35}-10^{36}$ erg/s. These are sufficient to explain the X-ray
luminosities of the Anomalous X-ray Pulsars, a class of neutron stars
whose X-ray luminosity is 1-2 orders of magnitudes higher than their
spin down energy, at odds with the more traditional radio pulsars,
whose X-ray emission is powered by their rotational energy loss.
Fallback disks have also been invoked (Xu et al. 2003; Shi \& Xu 2003)
to explain the observational properties of the central compact objects
(Pavlov et al.  2004) in supernova remnants.

The effects of fallback disks on the spin evolution of neutron stars
have been discussed by a number of authors. Marsden, Lingenfelter \&
Rothschild (2001) pointed out how the interaction of a putative
fallback disk with the magnetosphere of a NS can significantly modify
its spin down rate, and as such help cure the discrepancy between the
characteristic and the actual age that has been observed in some
pulsars. Menou et al. (2001a; see also Chen \& Li 2006; Yan et
al. 2012) showed how the extra torque provided by this putative
fallback disk can also result in braking indices which depart from the
value of 3 expected for spin down by pure dipole radiation. Blackman
\& Perna (2004) further noticed that the presence of such a disk can
play an important role for producing the jets that are observed in
some of these pulsars, such as Crab and Vela.  Qiao et al. (2003)
proposed a fallback disk model for periodic timing variations of
pulsars.  Cordes \& Shannon (2008) were able to explain the properties
of the recently discovered transient pulsars by means of asteroids
formed from a fallback disk and occasionally migrating into the pulsar
light cylinder.  The mass ejected by fallback in a supernova explosion
has been invoked as a production site for the r-process (Fryer at
al. 2006), as well as a potential source of power to SN light curves
(Piro \& Ott 2011; Dexter \& Kasen 2013).

From an observational point of view, fallback disks around NSs of
$10^4-10^5$ yr are expected to be especially bright in infrared
radiation (Perna, Hernquist \& Narayan 2000), and there have been many
searches aimed at detecting such emission (see e.g. Mignani et
al. 2007 for a review), resulting in some strong hints of possible
detection (Wang, Chakrabarty \& Kaplan 2006).

All the various models and scenarios discussed above, proposed to
explain a large variety of astrophysical phenomena, especially in the
context of the NS compact object remnants, depend {  crucially} on
the specific properties of the fallback disks.  In particular they depend on
the specific angular momentum content, which eventually
 determines the post-fallback circularization radius. Once fallback has
ended, this initial radius, coupled with the local viscous timescale,
will determine the initial accretion rate.  After their formation, the
evolution of fallback disks can be modeled following the results
of semi-analytical and numerical studies (e.g Cannizzo, Lee \& Goodman
1990; Cannizzo \& Gehrels 2009; Shen \& Matzner 2013): an early phase of constant accretion
rate, lasting on the order of the viscous time, followed by a powerlaw
decay. Once the initial accretion rate $\dot{M}_0$ is known, then the
subsequent evolution is determined in a self-similar fashion. Crucial,
therefore, are the {\em initial conditions}, i.e. the mass and
location of the matter that falls back and hence determines
$\dot{M_0}$. These can only be determined accurately by means of
numerical simulations of the collapse of massive rotating stars, and a
distribution of angular momenta.

Numerical simulations of pre-SN profiles in stars with a range of
masses and metallicities have been performed by a number of authors
(e.g. Woosley \& Weaver 1995; Fryer \& Heger 2000; MacFadyen, Woosley
\& Heger 2001; Heger, Woosley \& Fryer 2003; {  Woosley} \& Heger 2006;
Yoon, Langer \& Norman 2006; Zhang, Woosley \& Heger 2008; Woosley \&
Heger 2012). The aims of these works ranged from an investigation of
the properties of GRB progenitors, to a determination of the
rotational properties of the compact object remnants, to the
yields of heavy elements produced by nucleosynthesis. Among the
studies specifically devoted to fallback disk formation, the focus has
been on the compact, hyperaccreting disks believed to power GRBs.

\begin{figure*}
\begin{minipage}[t]{0.5\textwidth}
\centering
\includegraphics[width=8.2cm]{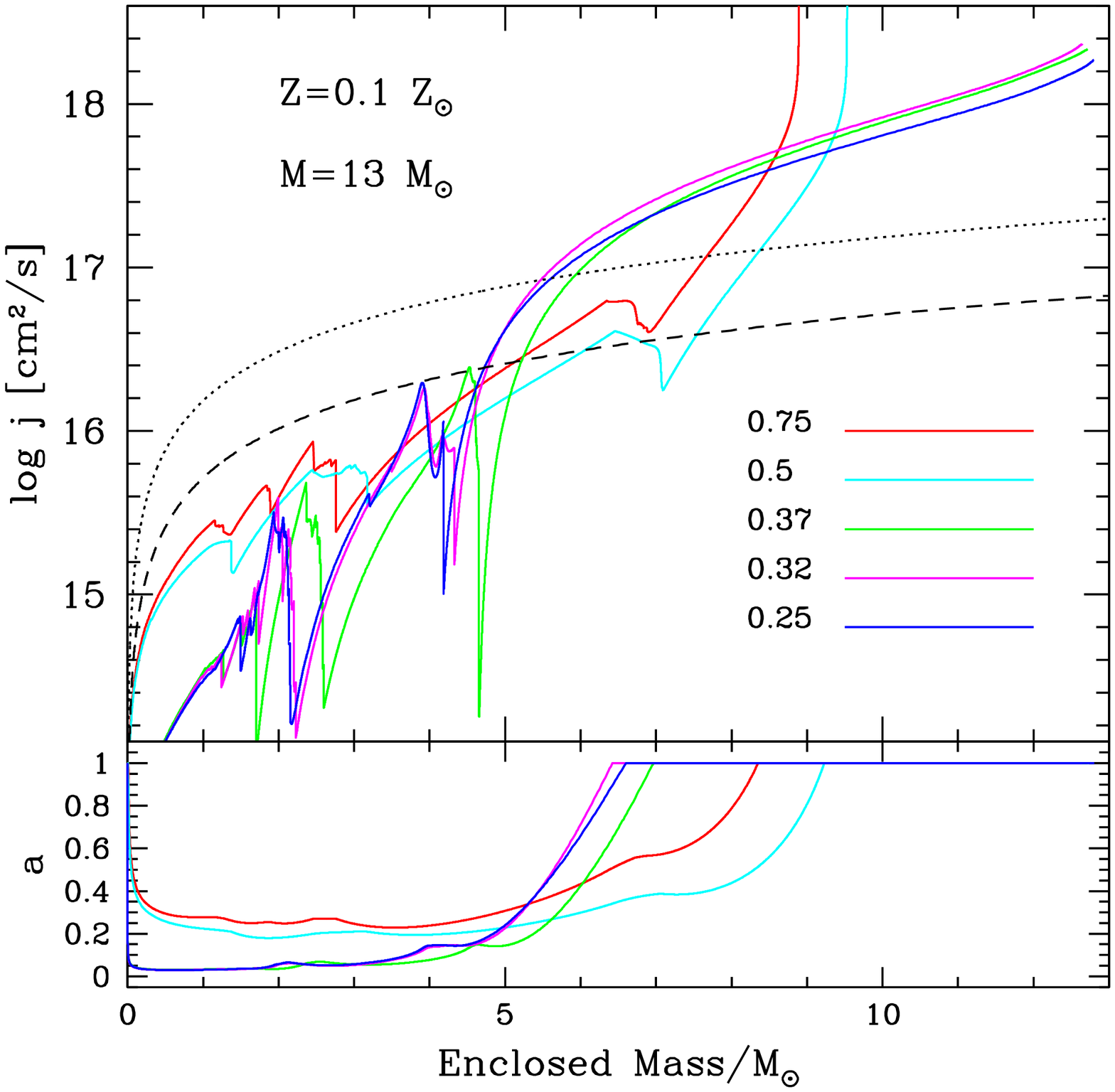}
\end{minipage}
\begin{minipage}[t]{0.5\textwidth}
\centering
\includegraphics[width=8.2cm]{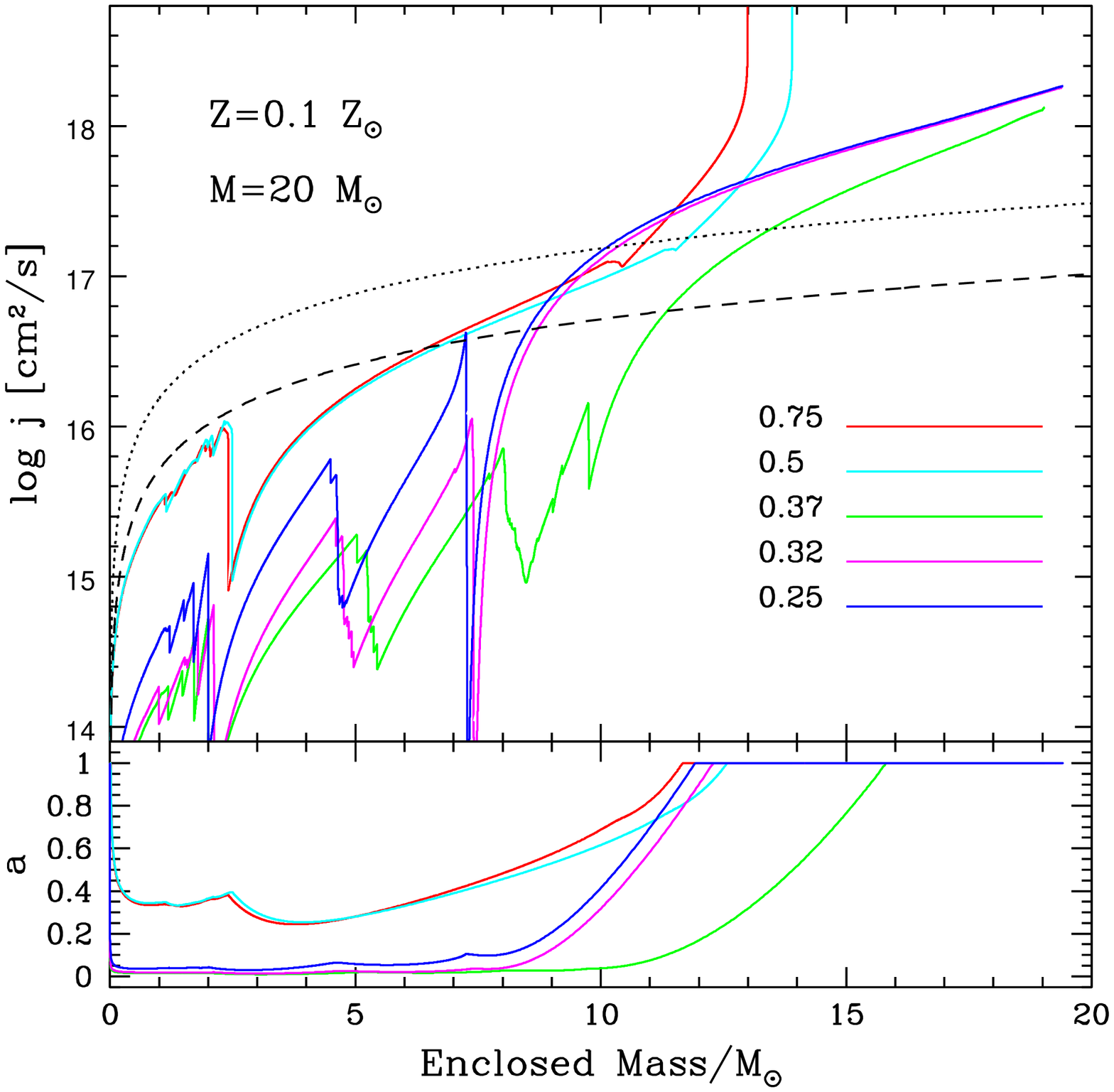}
\end{minipage}
\begin{minipage}[t]{0.5\textwidth}
\centering
\includegraphics[width=8.2cm]{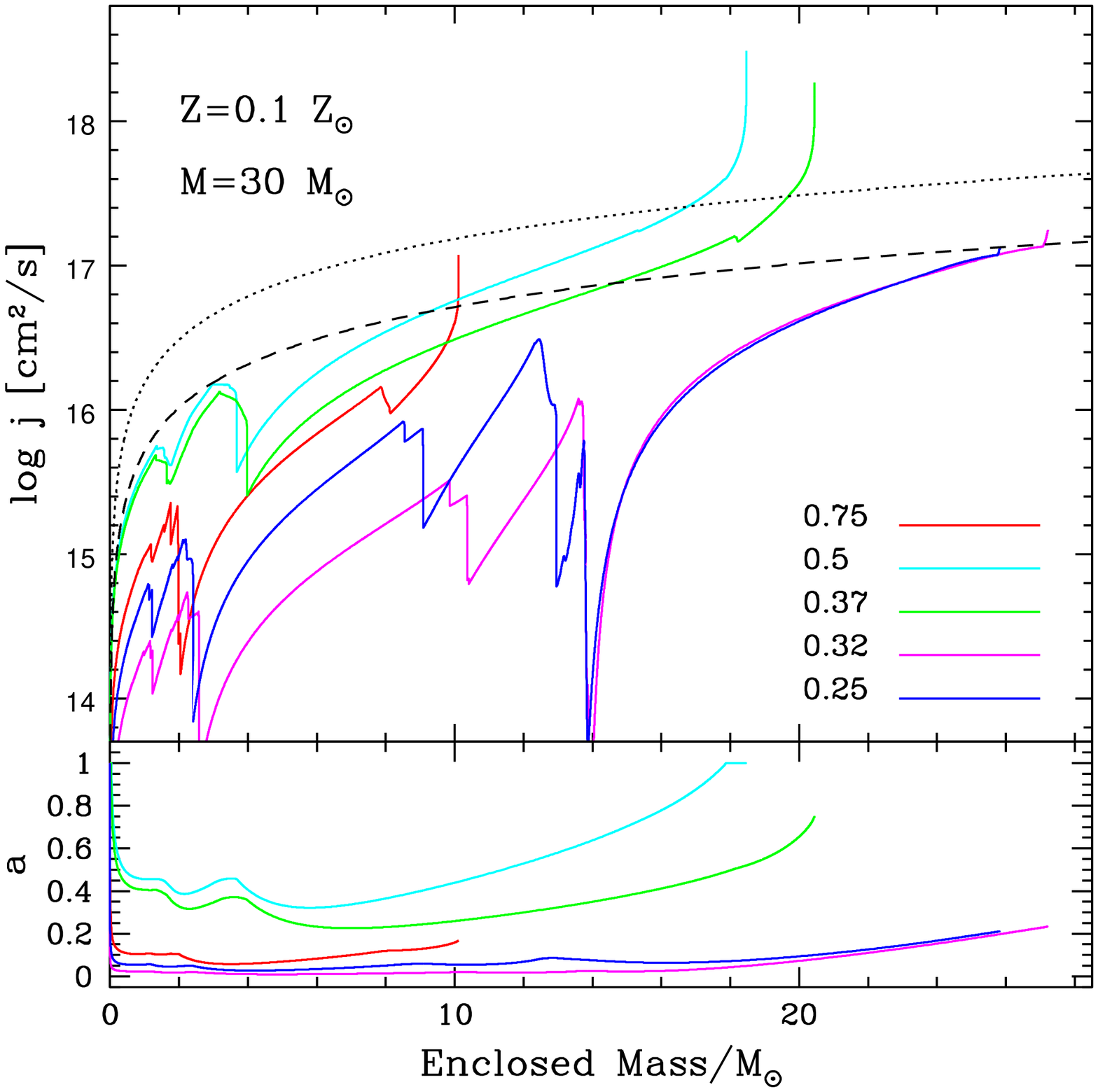}
\end{minipage}
\begin{minipage}[t]{0.5\textwidth}
\centering
\includegraphics[width=8.2cm]{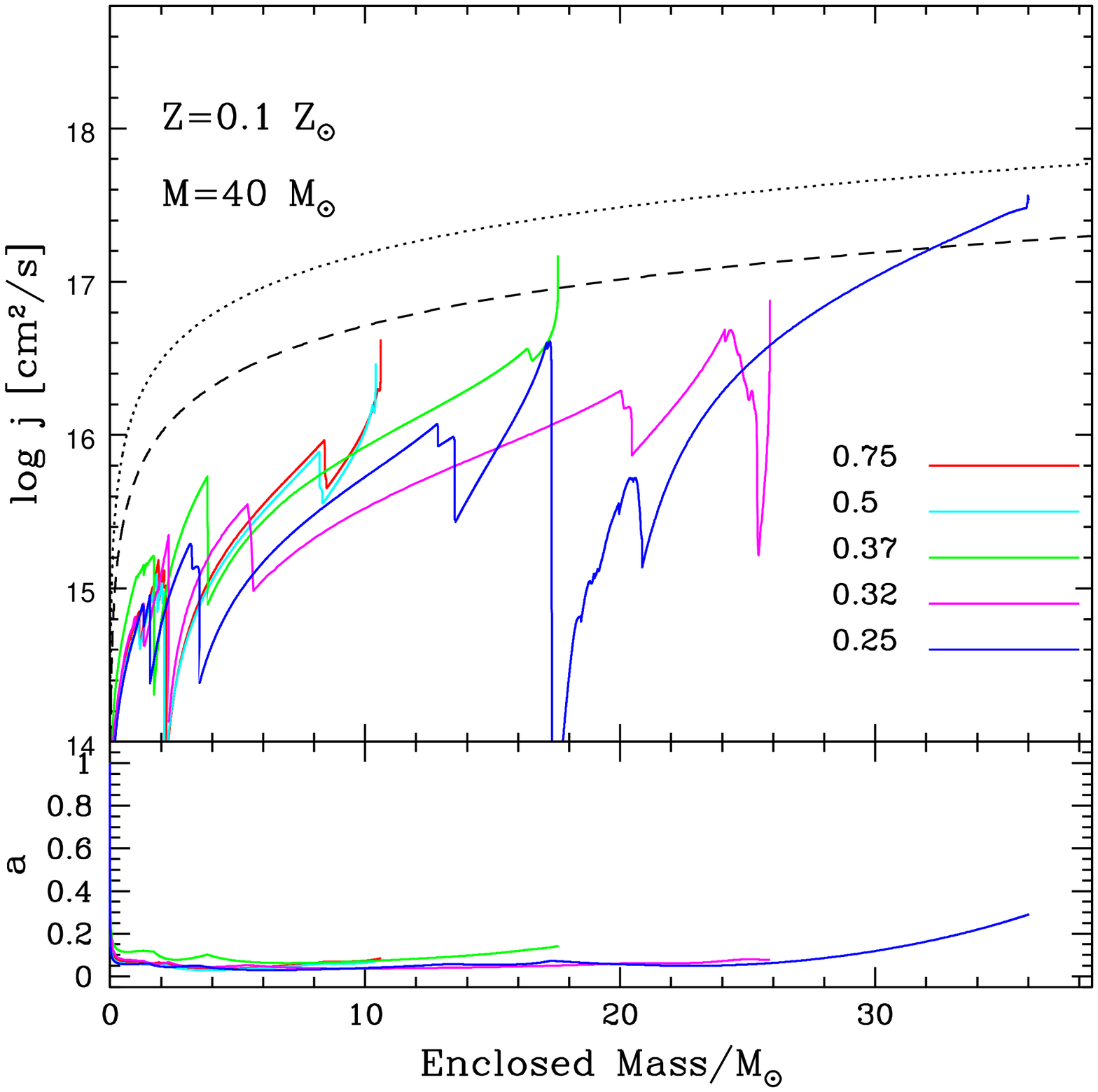}
\end{minipage}
\caption{Same as in Fig~\ref{fig:preSN1}, but for main sequence stars 
of 10\% solar metallicity.}
\label{fig:preSN2}
\end{figure*}

In this paper we take a new look at the problem of massive star
collapse, by concentrating our numerical study on the properties of
the fallback material that does not hyperaccrete rapidly onto the
newly born compact object, but that is instead able to form a
long-lasting accretion disk. We take advantage of the recent inclusion
of stellar rotation in the public MESA code for stellar evolution
(Paxton et al. 2013), which allows us to run a large grid of models.
Our study is aimed at determining the range of initial conditions
(mass, circularization radius) of the fallback material as a function
of mass and angular momentum of the progenitor star, both for the BH
and the NS remnant cases. For the special case of the NS remnants (not
treated so far in the numerical literature), we aim at assessing
whether the properties of these fallback disks around newly born NSs
are consistent with those needed to explain a wide range of NS
phenomena as described above.

Our paper is organized as follows: in Sec.~2 we describe the grid of
main sequence star models that we use (60 models, organized in a grid of 3
metallicities, 4 masses, and 5 initial angular velocities), and how
they are evolved with MESA.  Given the importance of magnetic torques
in the transfer of angular momentum between various layers, we further
explore, for a subset of cases, the dependence of the pre-SN angular
momentum on the presence (or not) of such magnetic torques.
Sec.~3 describes how the stars are exploded, and the resulting fallback mass
as a function of explosion energy and initial star configuration.
Sec.~4 takes the results from Sec.~2 and 3 in order to evaluate the properties
of the stars for which fallback disks are expected, and, when so, to
compute their initial properties and estimate their time evolution. 
We find that extended, long-lived disks around isolated NSs require
very fine-tuned initial conditions, and hence they are not expected
to be a common outcome of SN explosions. Extended, long-lived disks
could instead be found around BHs of $M\ga 10-15~M_\odot$, with conditions
possibly suited to planet formation.
We summarize our findings in Sec.~5.

\begin{figure*}
\begin{minipage}[t]{0.5\textwidth}
\centering
\includegraphics[width=8.2cm]{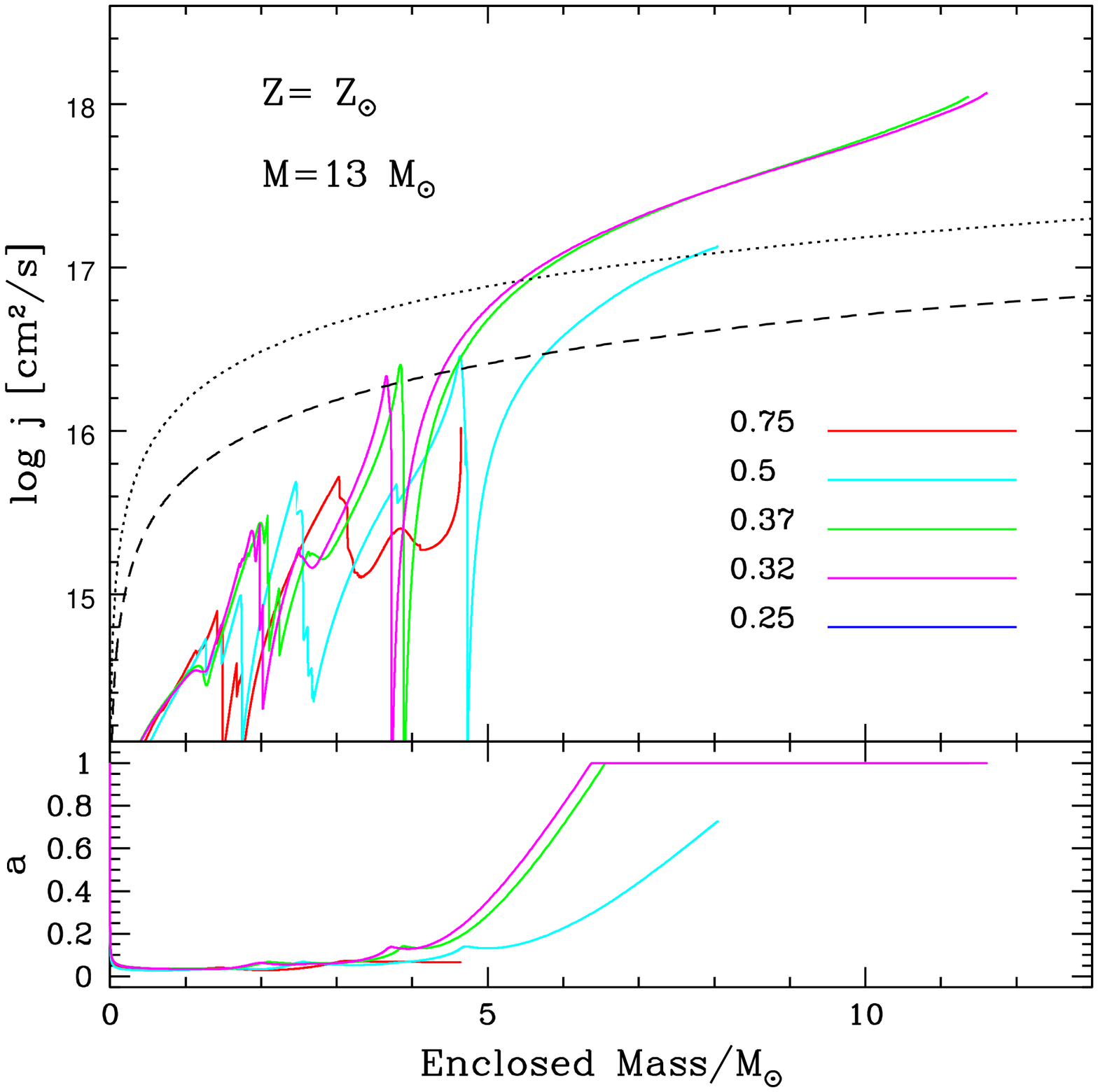}
\end{minipage}
\begin{minipage}[t]{0.5\textwidth}
\centering
\includegraphics[width=8.2cm]{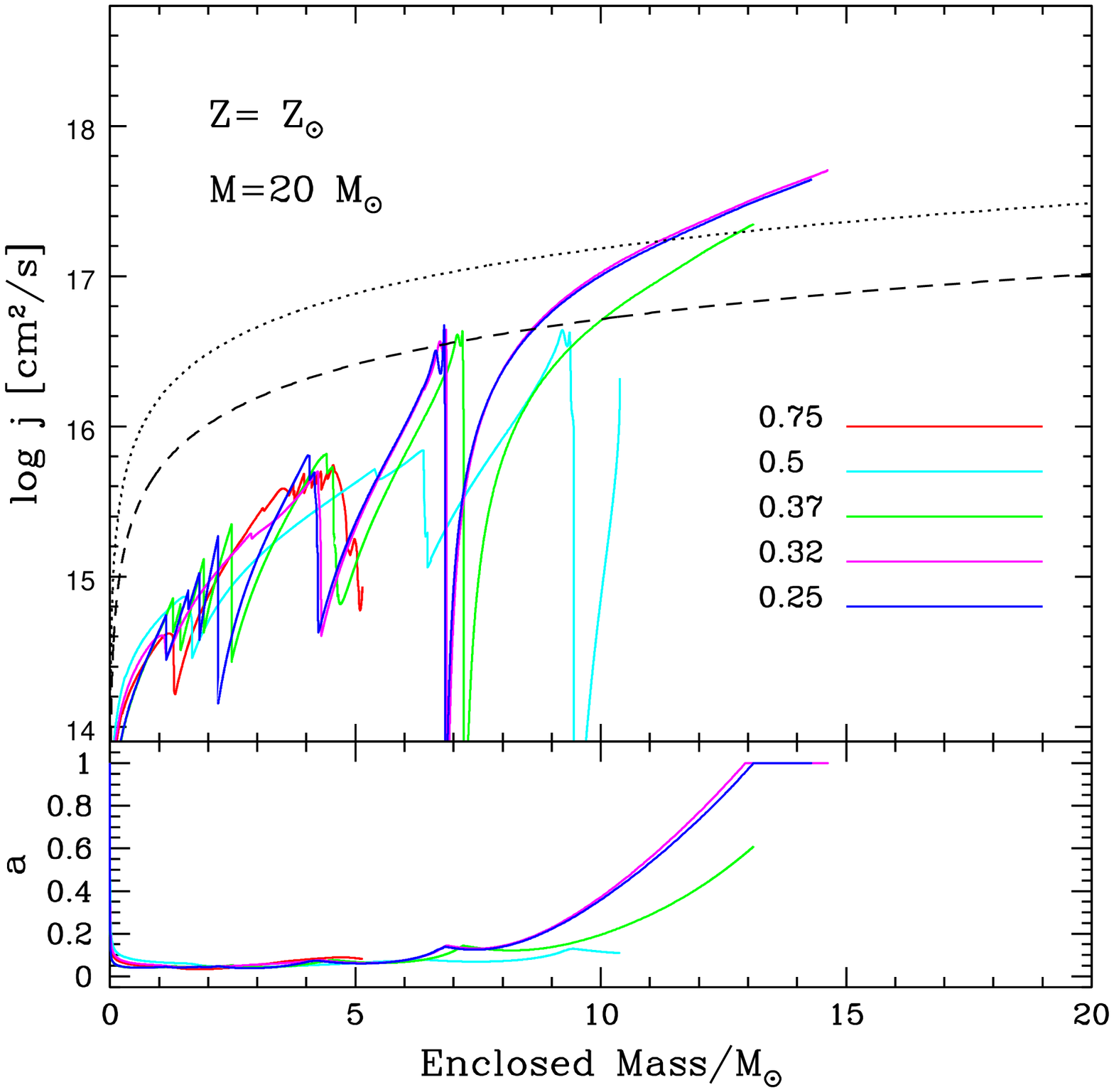}
\end{minipage}
\begin{minipage}[t]{0.5\textwidth}
\centering
\includegraphics[width=8.2cm]{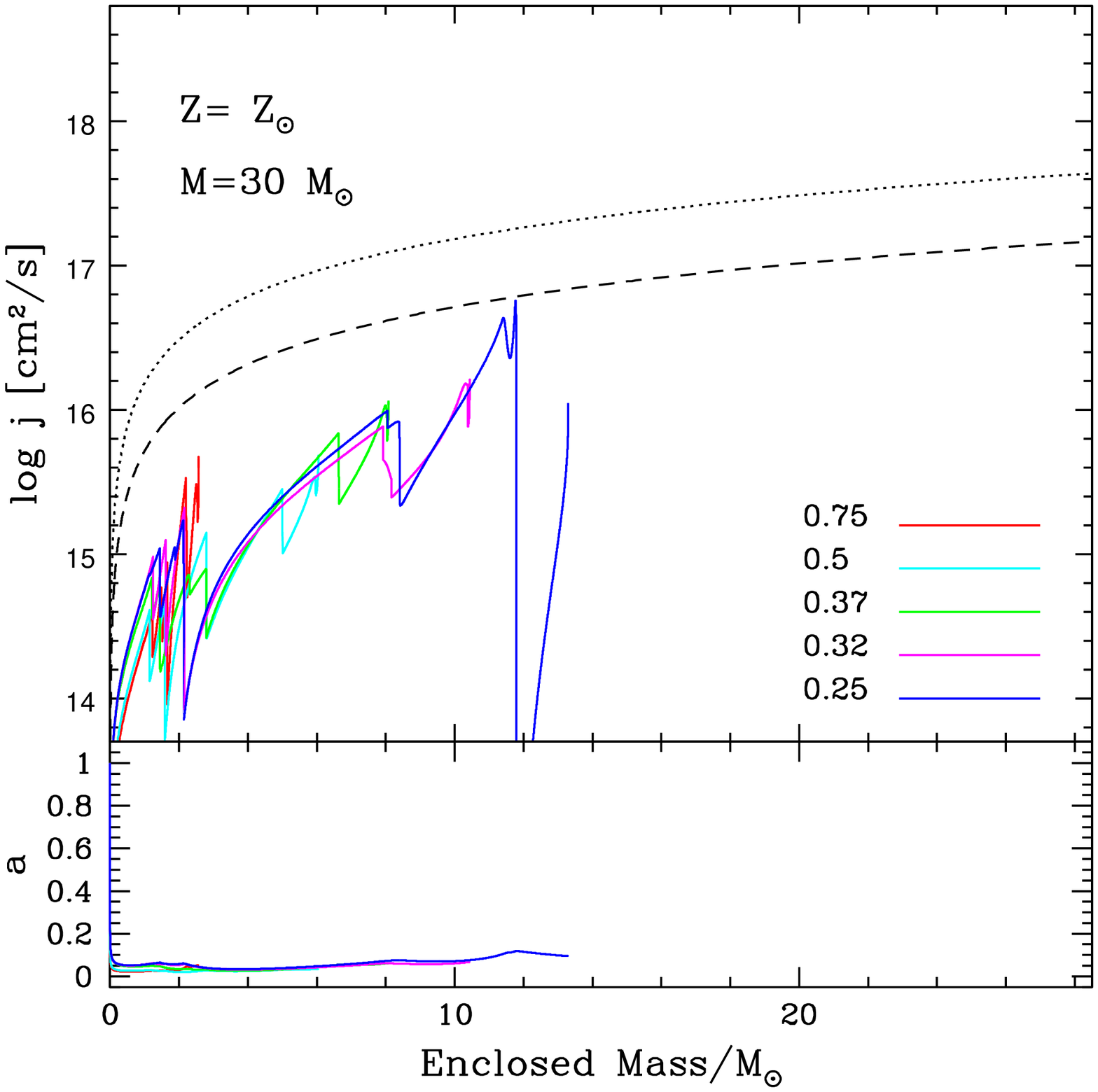}
\end{minipage}
\begin{minipage}[t]{0.5\textwidth}
\centering
\includegraphics[width=8.2cm]{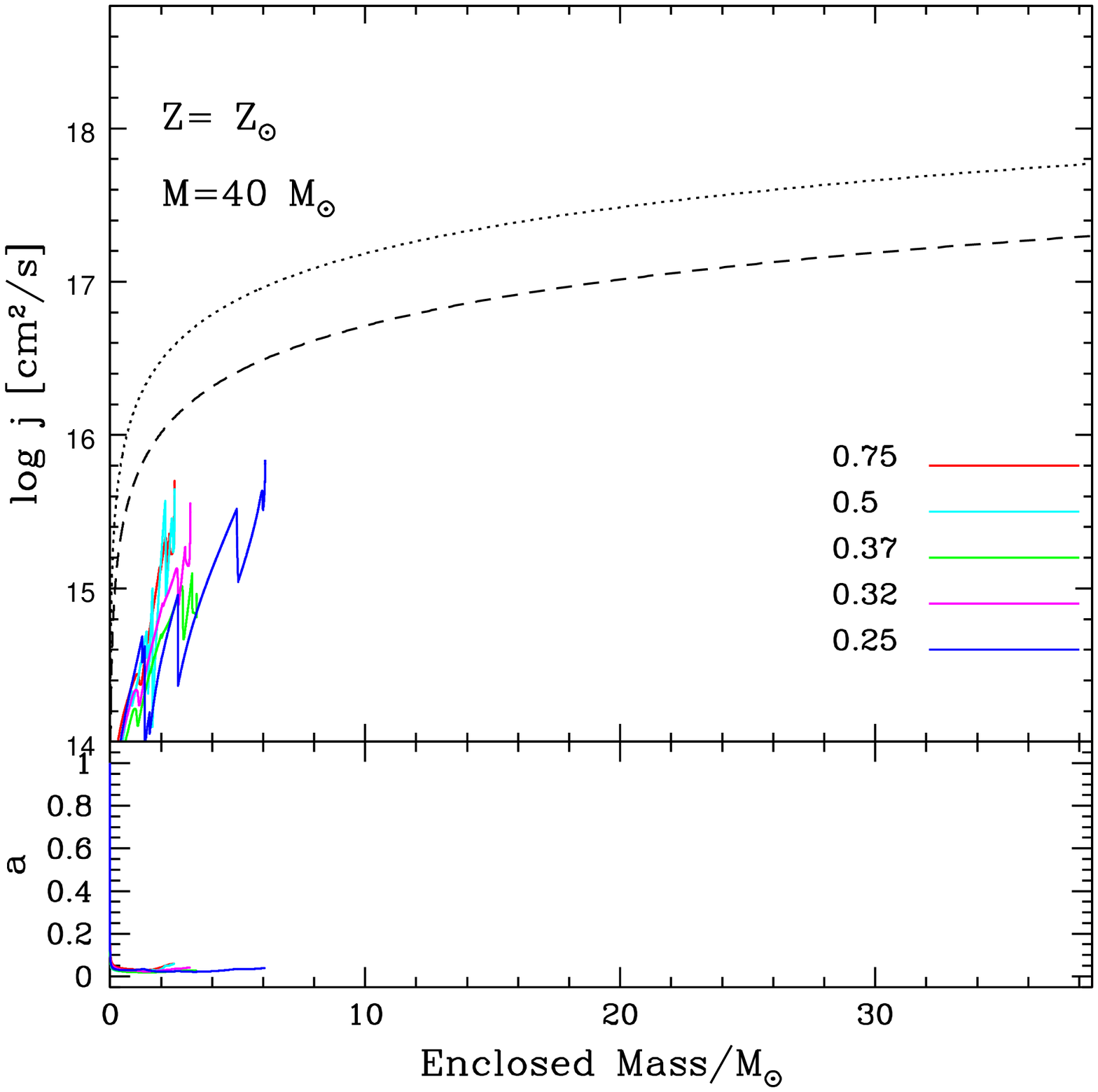}
\end{minipage}
\caption{Same as in Fig~\ref{fig:preSN1}, but for main sequence stars 
of solar metallicity.}
\label{fig:preSN3}
\end{figure*}

\section{Massive star models and their pre-supernova angular momentum distribution}

The fate of fallback matter after the SN explosion is primarily
determined by two factors: the angular momentum distribution of the
pre-SN star, and the explosion energy during the SN. The former
determines whether there is {\em any} matter with specific angular
momentum larger than the local Keplerian value, while the latter
determines whether that high-$j$ matter remains bound and hence can
circularize around the newly born compact object. We describe in this
section the computation of the pre-SN models, while in the following
section we describe how the stars are exploded.

We compute the pre-supernova angular momentum profiles by evolving
star models with the state-of-the-art open source code MESA
  (Paxton et al. 2011, 2013).  This code includes the physics of
  rotation, the implementation of which has been extensively discussed
  in Paxton et al. (2013). We refer to this latter work for the
  details of the angular momentum transport mechanisms due to rotation
  and magnetic torques adopted in this paper.

A number of studies (Maeder 1987; Yoon \& Langer 2005; Woosley \&
Heger 2006; Yoon et al. 2006; Paxton et al. 2013) have shown that, at
a certain threshold (surface) rotation rate $(\vsurfi/\vcrit)_{\rm thres}$, 
the evolution on the H-R diagram bifurcates. 
The critical surface equatorial velocity $\vcrit$ is defined as
\begin{equation}
\vcrit^2\; =\; \left( 1\,-\,\frac{L}{L_{\rm Edd}}\right)\,\frac{G\,M}{R}\,
\label{eq:vcrit}
\end{equation}
where $R$ is the radius of the star of mass $M$, and $L_{\rm Edd}=4\pi
cGM/\kappa$, with $\kappa$ the opacity.  Above this threshold,
internal mixing processes are able to inhibit the formation of a
strong compositional gradient in the star, which is otherwise built by
nuclear burning in the stellar core.  What ensues is a chemically
homogeneous evolution. Stars evolving this way avoid a core-envelope
structure and become compact Wolf-Rayet stars already during
H-burning. Their core, which is not dramatically spun down by coupling
with an extended envelope, retains a high angular momentum prior to
collapse if the wind mass-loss is not too strong (e.g. at low
metallicity).  On the other hand, stars below this threshold undergo a
``canonical'' evolution and experience a RSG phase, with their core
losing a substantial amount of angular momentum. The precise value of
$(\vsurfi/\vcrit)_{\rm thres}$ is firstly a function of the mass of
the star, with a further dependence on metallicity.  More massive
stars evolve chemically homogeneously at a lower value of
$(\vsurfi/\vcrit)$. This is because the Eddington-Sweet circulation,
which, beside convection, is believed to be the dominant mixing
process during the main sequence of rotating massive stars, has a
timescale proportional to the Kelvin-Helmholtz time scale,  $\tau_{\rm ES}\propto \tau_{\rm KH} \,(\vsurfi/\vcrit)^{-2}$. For main
sequence stars the Kelvin-Helmholtz timescale decreases with
increasing mass.  Moreover, in massive stars the increasing radiation
pressure weakens the entropy barrier, resulting in more efficient
internal mixing (Yoon et al. 2006).  On the other hand, higher
metallicity means an increasing amount of mass loss (and hence angular
momentum loss) through line-driven winds. This tends to slow the star
down, requiring a larger value of $(\vsurfi/\vcrit)$ for the star to
follow the chemically homogeneous branch.  Note that even if a star
begins evolving chemically homogeneously on the zero age main
sequence, its wind can quickly remove angular momentum and the star
moves back to a canonical evolution. In our calculations mass
loss has been implemented according to the recipe described in Yoon et
al. (2006). Note that massive rotating stars can approach critical
rotation during their evolution, which is expected to lead to an
enhancement of the stellar mass loss. This is calculated according to
the prescription (see Sec.~6.4 in Paxton et al. 2013)
\begin{equation}\label{mdotomega}
  \Mdot\left(\Om\right) = \Mdot(0)\,\left(\frac{1}{1-\Om/\Omc}\right)^{0.43},
\end{equation} 
where $\Om$ is the surface angular velocity, and $\Omc$ is the critical angular velocity 
at the surface ($\Omc^2=(1-L/\Ledd)\, GM/R^3 =\vcrit/R_{\rm crit}$, 
where $R_{\rm crit}$ is the equatorial radius at breakup, which is 1.5 times the polar radius).

\begin{figure*}
\begin{minipage}[t]{0.5\textwidth}
\centering
\includegraphics[width=8cm]{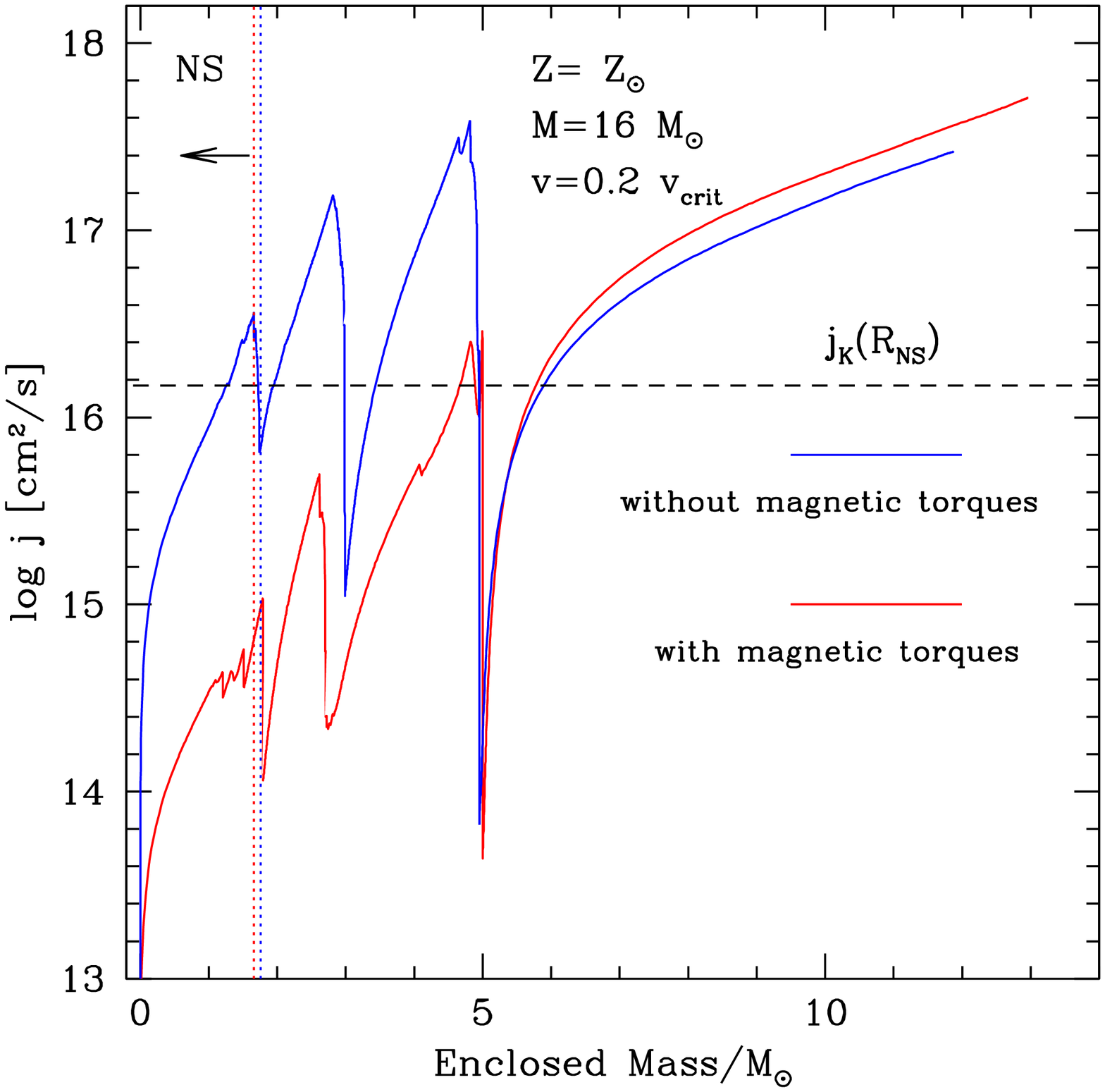}
\end{minipage}
\begin{minipage}[t]{0.5\textwidth}
\centering
\includegraphics[width=8cm]{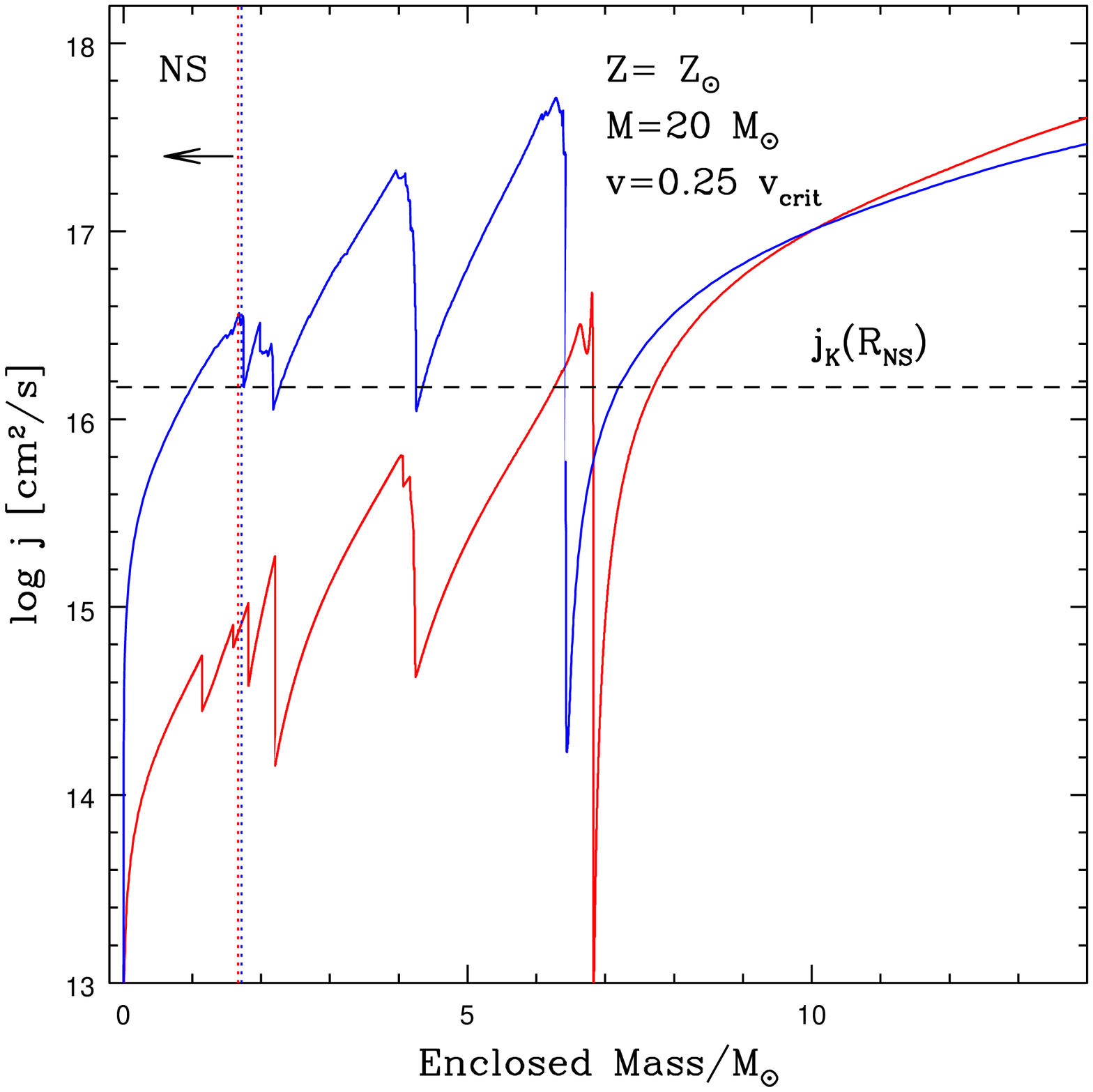}
\end{minipage}
\caption{Comparison between the distributions of specific angular
  momentum in the pre-SN star with and without the inclusion of
  magnetic torques.  These stars, at solar metallicity, leave behind a
  NS remnant.  The dashed horizontal line indicates the specific
  angular momentum needed for circularization at the NS radius, for a
  typical $R=10$~km.  The region to the left of the vertical lines
  indicates the mass which collapses to a NS.}
\label{fig:preSN-NOB}
\end{figure*}

The mass loss timescale is limited to the thermal timescale of the star $\tkh$
\begin{equation}\label{mdotomegatkh}
  \Mdot= \textrm{min}\,\left[ \Mdot(\Om) \,, 0.3\,\frac{M}{\tkh} \right]\,.
\end{equation}
Convective regions are determined according to the Ledoux
criterion. Semiconvection according to the prescription of Langer et
al. (1983) is used, with a parameter $\alphasc=1.00$.

In order to explore a significant parameter space in velocity, mass,
and metallicity, we use MESA to evolve a grid of 60 models,
characterized by 3 values of metallicity, ($Z=1\%, 10\%$ and $100\%\,
Z_\odot$), 4 values of mass ($M=13, 20, 30, 40\,M_\odot$), and 5
values of surface rotation velocity ($\vsurfi=0.25, 0.32,
0.37, 0.5, 0.75$ $\vcrit$).  As our ``standard'' set of models, we
consider calculations that include internal angular momentum transport
by magnetic torques. These magnetic fields arise in radiative zones
and are believed to be produced by dynamo action transforming part
of the rotational energy into magnetic energy (Spruit-Tayler dynamo;
Spruit 2002). While the reality of this dynamo loop in stars is
currently debated (e.g., Braithwaite 2006; Zahn et al. 2007), models
including this extra angular momentum transport mechanism do a much
better job predicting the final spin rate of compact remnants (Heger
et al. 2005; Suijs et al. 2008). Therefore, we have adopted here as
our standard evolution one that includes magnetic coupling. However,
for a few significant cases, we also explore the influence on the
results of the lack of magnetic fields during evolution.  Whenever we
need to refer to a specific model within the standard set, we will use
the notation ZxxMyyvzz, where xx is the percentage of solar
metallicity, yy is the mass in solar units, and zz is the percentage
of the critical velocity.

The MESA runs are set up to proceed until the onset of core collapse,
defined as the moment when any part of the Fe core is falling
  with a velocity $\varv \ge1000\,\kms$; however, for some models, the
timestep adopted by MESA past core Carbon burning become prohibitively
small. As at this time the star has only a few years to live, the
amount (and distribution) of angular momentum is
not expected to change substantially.  Therefore the conclusions that
we draw on the angular momentum distribution of the fallback matter
are expected to remain robust, regardless of the particular endpoint,
past core C-burning, chosen for the model.

{  Models at solar metallicity have been evolved using an artificial
  reduction of the superadiabaticity ($\supernab \equiv \nablaT -
  \nablaad$) in radiation dominated convection zones. Here $\nablaT$
  and $\nablaad$ are the actual temperature gradient and the adiabatic
  temperature gradient, respectively.  This reduction is applied in
  regions where the mixing length theory (MLT) is out of its domain of
  applicability and MESA (like other 1D stellar evolution codes)
  struggles with very short timesteps. This approach (MLT++) is
  thoroughly described in Paxton et al. (2013). In this study we reduced
  the superadiabaticity of $Z=Z_\odot$ models by a factor
$10^{-2}$  when $\supernab >  10^{-4}$.} 

\begin{figure}
\centering
\includegraphics[width=8cm]{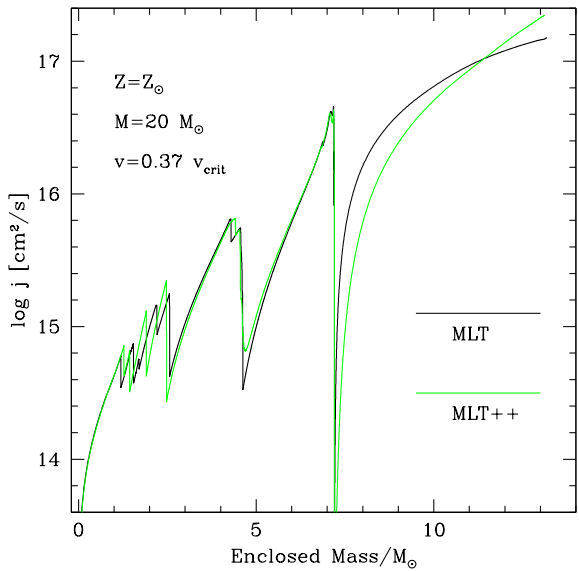}
\caption{  Comparison between the distributions of specific angular
  momentum in the pre-SN star with and without the reduction of
  superadiabaticity in radiation dominated envelopes (MLT++).  
The comparison is for a 20$M_\odot$ model at
  $Z=Z_\odot$ initially rotating at 37\% of critical velocity.}
\label{fig:preSN-MLT}
\end{figure}

Figures~\ref{fig:preSN1}, \ref{fig:preSN2}, \ref{fig:preSN3} show the
specific angular momentum distribution in the pre-SN stars for three
values of the metallicity ($0.01Z_\odot$, $0.1Z_\odot$ and $Z_\odot$),
and a range of ZAMS rotational velocities between 0.25 and 0.75 of the
critical value. For each of the models, we
also show the distribution of the spin parameter $a=J/M$ of the
enclosed mass.

While the above are our ``standard'' model, we also consider, for a
few cases, how the pre-SN angular momentum distribution is modified in
the case of no magnetic coupling.  Since this is of particular
relevance for the possibility of forming fallback disks around NS
remnants, we consider models which leave behind a NS.
Figure~\ref{fig:preSN-NOB} shows a comparison between models with and
without $B$ fields, for two {   solar metallicity models with mass $16$ and
20 $M_\odot$ and a rotational  velocity of  $\vsurfi/\vcrit=0.2$ and  $\vsurfi/\vcrit=0.25$}.  Also
indicated is the minimum specific angular momentum for material to
circularize just outside of a NS of mass equal to that of the iron
core in the simulation, and a typical radius of 10~km.  It is evident
how the lack of magnetic torques results in a substantially larger
specific angular momentum in the layers outside of the iron core. This
is because the core of the star evolves essentially decoupled from the
envelope, and is able to retain a large fraction of its initial
angular momentum.

{  To assess the effect of the adoption of the MLT++ in our solar
  metallicity models, we evolved a 20$\Msun$ model initially rotating
  at 37\% of its critical velocity, with and without the reduction of
  the superadiabaticity. A comparison of the final angular momentum
  distributions is shown in Fig.~\ref{fig:preSN-MLT}, revealing
  minimal differences between the two calculations.  For models above
  $\sim 25\Msun$ at solar metallicity, similar tests cannot be
  performed, as in that case calculations that do not include the
  MLT++ cannot reach core collapse (this is the reason why MLT++ has
  been implemented in the first place, see Paxton et al. 2013). In
  general, we expect the integrated mass and angular momentum loss to
  be affected by the adoption of MLT++, since this tends to change the
  surface stellar radius (hence surface temperature, entering the
  calculations of the mass loss rate). Such uncertainty is even larger
  for rapidly rotating models, where the evolution of the stellar
  radius (affected by the efficiency of energy transport in the
  stellar envelope) has an important role in determining the amount of
  mass lost at critical rotation. Until a better theory for energy
  transport in radiation dominated zones is available, we have to
  consider these results highly uncertain. Since the models at solar
  metallicity have not been considered for the explosion calculations
  (disk formation is marginal for $Z=Z_\odot$), the adoption of MLT++
  does not affect the stellar explosions calculations presented in
  Sec.~3 and, overall, the major conclusions of this paper.}

The fate of the material ejected during the explosion depends on
both its specific angular momentum, as well as on the energy of
the explosion, which determines the amount of mass that remains bound,
and hence falls back. The latter is discussed in the following section.

\begin{figure*}[t!]
\begin{minipage}[t]{0.5\textwidth}
\includegraphics[width=8.7cm]{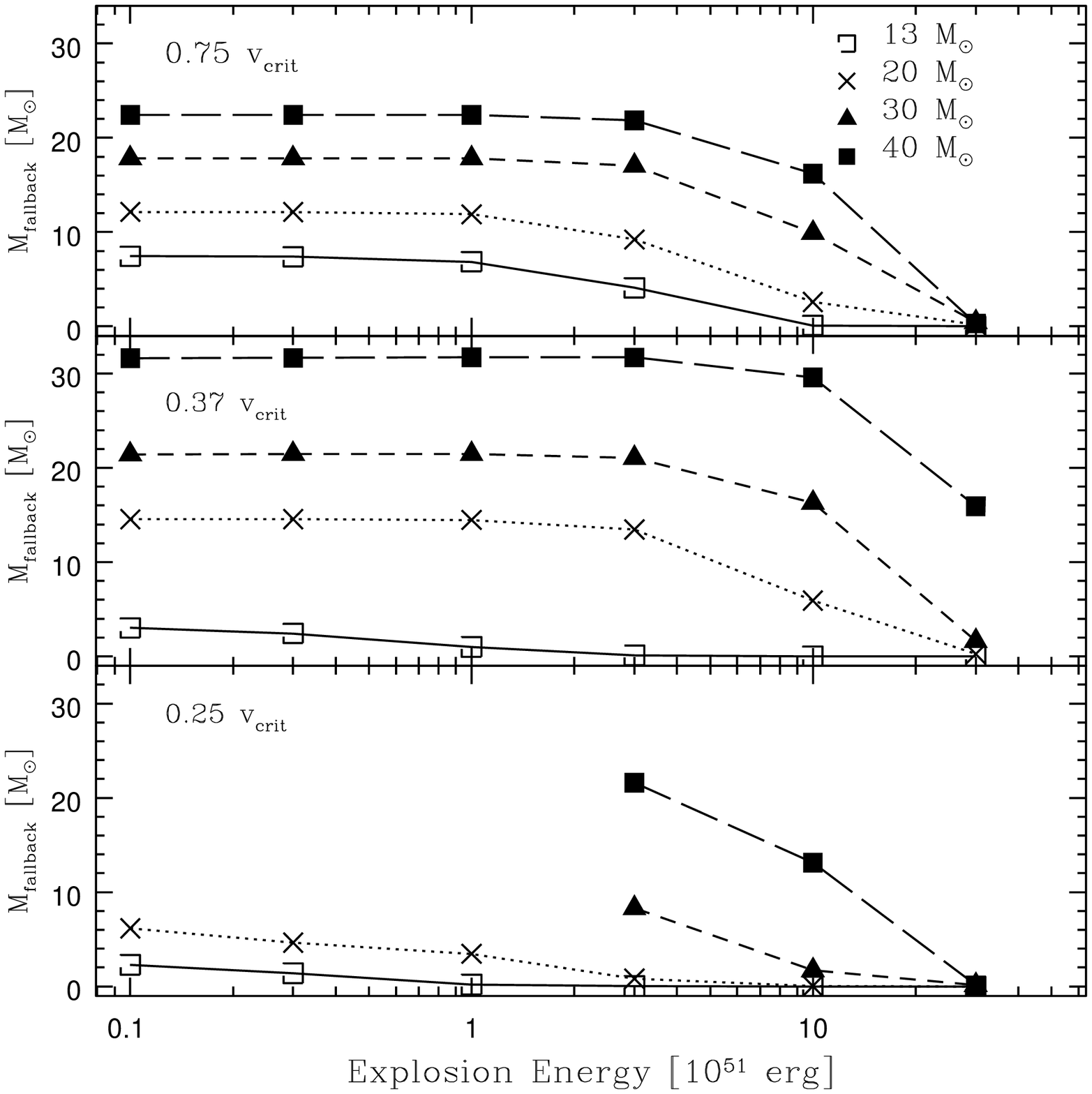}
\end{minipage}
\begin{minipage}[t]{0.5\textwidth}
\includegraphics[width=8.7cm]{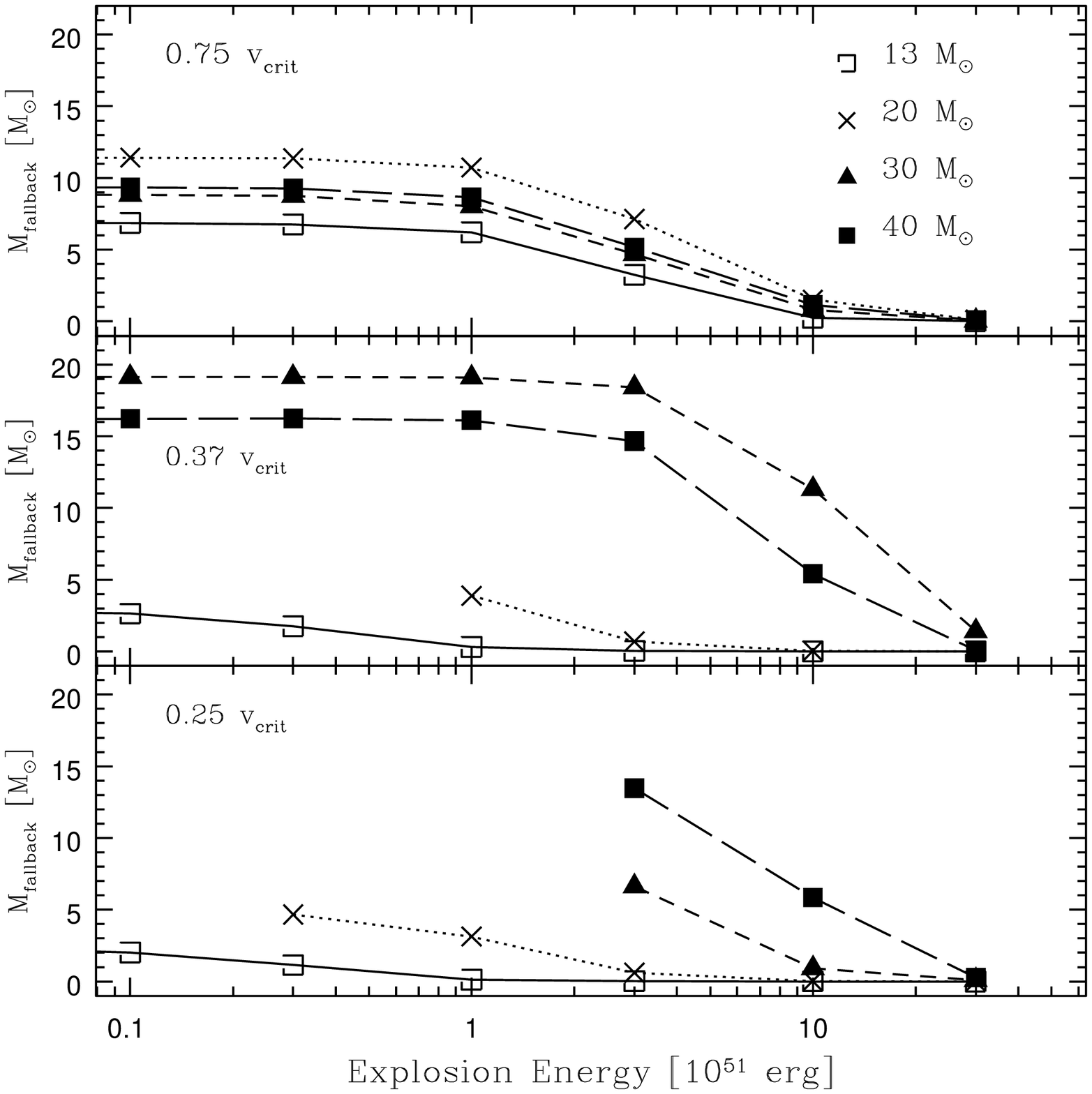}
\end{minipage}
\caption{Fallback matter (i.e. {  amount of mass which remains bound}) 
as a function of the explosion energy for
  stars with metallicity $Z=0.01 Z_\odot$ (left panel) and $Z=0.1
  Z_\odot$ (right panel), and for the 'standard model' with magnetic
  torques.  Results for three values of the initial rotational
  velocity of the star are displayed. We have not displayed the data points
  for those simulations for which the shock had not had yet the time to
  cross the outer layers of the star during the run time of the
  simulation. This is the case for those stars which do not undergo
significant mass loss during their evolution, and hence have extended
envelopes in the pre-SN state. }
\label{fig:fallback}
\end{figure*}

\section{Stellar explosions and their fallback mass}
The fallback dynamics are computed by means of direct numerical
calculations of the explosion.  These calculations are performed using
a one-dimensional version of the TESS code (Duffell \& MacFadyen 2011,
2013). TESS is a multi-dimensional moving-mesh hydrodynamics method,
which is both Lagrangian and shock-capturing.  As a result, the
expansion of the shock through the many layers of the star is captured
very accurately.

Each calculation is initiated as follows: The final state of a MESA
run is used as initial conditions for the density and pressure of the
star.  The mass of the iron core is subtracted from this density
profile and re-introduced as a point mass, to simulate the formation
of the compact object.  A fixed thermal energy is then added within a
small radius (1.5 times the radius of the core), which causes an
expansion and accelerates a shockwave which pushes its way out of the
star.  The inner boundary is initially reflecting for the first phase
of evolution, as the gas is accelerating. After this thermal energy
has been converted into kinetic energy in the shockwave, fallback can
begin (this typically takes about 10 seconds). At this point, we
switch the boundary to be absorbing, and mass can freely pass through
the inner boundary and be added to the compact point mass at the
origin.  After the shock has crossed a given radius r, fluid elements
interior to this radius have been accelerated and can either escape
the gravitational potential of the interior mass, or freely
fall back.  At any given time, we can define a critical radius,
$r_{\rm bound}$, as the largest radius such that all material interior
to it is bound.  We calculate this radius and measure the amount of
mass in its interior, $M_{\rm bound}(t)$, as a function of time.  At
late times, this mass asymptotes to a finite value $M_{\rm
  bound}(\infty)\equiv M_{\rm fallback}$, which is the total fallback
mass. 

{  For each pre-SN model, we simulated the star explosion for a
  broad range of explosion energies. We note that the mechanism for
  core collapse supernova explosions remains a topic of current
  intensive research. Indeed, many detailed numerical simulations
  (e.g. Herant et al. 1994; Burrows et al 1995; Woosley \& Janka 2005;
  review by Janka et al. 2007 and references therein) are not able to
  achieve a successful explosion, especially for the higher mass
  stars. Since the explosion dynamics depends on complex non-linear
  interactions between neutrinos interacting with fluid in a turbulent
  ``gain'' region (e.g. Liebendorfer et al. 2005), it is still not
  possible at present to calculate the explosion energy with any
  certainty.  In the future, high resolution 3D simulations with full
  neutrino transport schemes may be able to solve this
  problem. Lacking a definitive detailed calculation for the explosion
  dynamics, the precise kinetic energy released must be constrained by
  observations.  Here we have used a range of kinetic energies
  consistent with observations as well as with the recent literature
  (e.g. Nomoto et al. 2006; Mazzali et al. 2013).}

Fallback masses for a range of star models and explosion energies are
plotted in Figure~\ref{fig:fallback}; in particular, for the standard
models with magnetic coupling, we simulated the explosion of stars
with metallicities of $1\%$ and $10\%$ of the solar value, since the
possible formation of a disk is marginal at solar metallicity in these
models.

Note that, due to computational limitations, we do not explicitly
evolve the system to all the way to fallback, as this process can take an arbitrarily
large amount of time for marginally bound fluid elements.  Typically
we evolve the system for a time $t_{\rm fin} = 2\times10^3$~seconds.  We
should further note that, for some models, this running time is not
even enough for the shock to push its way through the entire star.  In
particular, models with extended envelopes which have not undergone
substantial mass loss take a very long time for the shock to cross the star
($\sim 10^7$~seconds). This was generally found to be the case in the
models with slow rotation ($\vsurfi = 0.25 \vcrit$), and also for the
$10\%$ metallicity example at moderate rotation ($\vsurfi = 0.37 \vcrit$).
Longer run-times are computationally expensive, especially given the
large number of models and explosion energies being calculated here.
However, for these cases, the outer layers of gas have a relatively
low binding energy, and hence even a very weak explosion can unbind this
mass.  If the explosion energy is much larger than this binding
energy, we expect the results to not depend sensitively on the outer
layers.  For a few of the models (in particular Z1M13v32 and
Z10M13v32, see Sec.~4.2.2), we have explicitly run the
calculation to a much longer time, $t_{\rm fin} = 10^7$~seconds, to verify
that these outer layers do indeed behave as we expect them to.

For the reasons mentioned above, we have found in practice that the
details of these marginally-bound outer layers are only important for
very weak explosions.  For these models, we have omitted the
low-energy results from Figure~\ref{fig:fallback}, so that the only
data included represent simulations for which either the shock has
crossed the star, or the outer layers do not significantly affect
explosions of the given energy.

\section{Post-SN disk formation, evolution, and observability}

In the following, we couple the results from the previous two sections
to discuss the post-SN phenomenology of the ejected material. We treat
the NS and BH cases separately.

For each star model, in order to establish whether the iron core
immediately collapses to a NS or a BH, we track its mass. For the
models we studied, masses are below $\sim 2.1 M_\odot$ in all the
cases. A NS is expected to be the common outcome of the implosion of
the iron core immediately after the SN explosion\footnote{These
  results are consistent with the findings of Heger et
  al. (2003). They found that only for initial star masses $M\ga 40
  M_\odot$ (and sub-solar metallicity) a BH can be formed by direct
  collapse.}.  However, for a large range of explosion energies
enough material is expected to fall back and form a BH. BH formation
from fallback has been widely discussed in the literature (e.g. Heger
et al. 2003). In the following, for both NSs and BHs, we will focus 
our discussion on the angular momentum of the material that falls
back, and hence on the possibility of the existence of a short or
long-lived fallback disk.

For a fallback disk to be formed, a fraction of material must remain bound,
and it must possess a specific angular momentum $j_{\rm m}$ at least as large
as the angular momentum at the last stable orbit, $j_{\rm ls}$. 

The bound fraction of the stellar envelope at radius $R$ will fall back on a timescale 
on the order of the free-fall time
\begin{equation}
t_{ff} (r)= \pi \sqrt{\frac{R^3}{8 G M}}\,,
\label{eq:tff}
\end{equation}
where $M$ is the enclosed mass inside radius $r$. The maximum radial
distance, $r_{\rm max}$, reached by each parcel of material is determined by the details
of the explosion, with the explosion energy playing the most important role. 
We compute the post-explosion density profile of the gas as described
in Sec.~3. 

Each bound parcel of gas, of specific angular momentum $j_{\rm m}$,
after falling back will circularize at a radius $R_{\rm circ}$ given
by the solution of the equation $j(R_{\rm circ}) = j_{\rm m}$.  Once
the bound material has fully circularized in a ring, the following
evolution occurs on the viscous timescale\footnote{  We note that,
  while Eq.~(\ref{eq:tvisc}) represents the classical viscous timescale,
  material can flow at a faster rate if transition fronts (see
  e.g. Menou et al 1999) are present (Cannizzo 1998; Kotko \& Lasota
  2012). }
\begin{equation}
t_0\,(R_{\rm circ})\,=\,\frac{R_{\rm circ}^2}{H^2 \alpha \Omega_K}\,
\sim 275\; \alpha^{-1}_{-1}\;m^{-1/2}_{10}\,
R_{10}^{3/2}\left(\frac{R}{H}\right)^{2}\,{\rm s} \;,
\label{eq:tvisc}
\end{equation}
where $m_{10}=M/(10~M_\odot)$, $R_{10}=R/(10^{10}~{\rm cm})$, $\Omega_K$ is the Keplerian
velocity of the gas in the disk, $H$ the disk scale-height, and
$\alpha$ the viscosity parameter (Shakura \& Sunyaev 1973), written in
units of $\alpha_{-1}\equiv\alpha/0.1$. 

The accretion rate at early times, while fallback still proceeds, is
determined by the longer of the two timescales above. The subsequent
evolution of the material, and its observable phenomenology, will depend
on the initial $\dot{M}$, as well as on the location at which the fallback
ring of material circularizes. 

In the following, we will discuss the expected initial conditions of
the high-$j$ fallback material for the wide range of progenitor stars 
and explosion energies considered here.

\subsection{Fallback and disk formation around Neutron Stars}

We begin with the discussion of the expectations for our standard models
with magnetic torque. The remnant compact objects are NSs 
for strong explosions, with the precise value of $E$ depending on
$M,Z,v$.  

Inspection of Figs.~\ref{fig:preSN1},
\ref{fig:preSN2} and \ref{fig:preSN3} shows that, if the stellar
evolution proceeds under conditions of strong magnetic coupling, the
outermost shells of material surrounding the newly born NS {\em never}
possess sufficient angular momentum to be able to circularize into a
disk, even for the fastest rotating main sequence stars at low
metallicity. The circularization radius for the small shells of
material just outside the NS is at most a couple of kilometers for the most
favorable scenarios.  In these situations, if those shells are not
ejected during the explosion, that material will fall back and impinge
on the NS surface before completing a full orbit. These sudden episodes
of accretion might give rise to X-ray flashes immediately following
the NS birth. However, the high opacity of the stellar envelope at those
early times (e.g. Chevalier \& Fransson 1994) will make these re-brightenings
difficult to observe.

The fate of the fallback material after the SN explosion is expected
to be very different if magnetic torques do not play any significant
role during stellar evolution.  To explore this scenario, we ran two
new sets of models; each set has the same $Z, M, v$, but one case is
with magnetic torques, and the other without. The comparative results
are displayed in Fig.\ref{fig:preSN-NOB}, for stars of solar
metallicity. 
It is apparent that, lacking magnetic torques to slow down the
interior layers, there is plenty of material with specific angular
momentum large enough to circularize outside of the NS
radius. However, the explosion energy needs to be rather fine-tuned so
that not all the outer layers are ejected, nor too much of them falls
back causing prompt collapse to a BH.

Let's consider the case that the explosion energy happens to be just
about right so that only a small fraction, say $\la 0.1 M_\odot$,
falls back. With the initial collapsing core $\sim 1.7 M_\odot$ in the
considered model, such a fallback mass amount ensures that the compact
remnant remains an NS.  We exploded the Z100M16v20 model shown in
Fig.~\ref{fig:preSN-NOB} (left panel, profile without magnetic
torques) and found that, for an explosion energy $\sim 3\times
10^{52}$~erg, about $0.08~M_\odot$ of solar mass remains bound, while
the rest of the envelope gets unbound. This bound $\sim 0.08~M_\odot$
is located at a radius of about $2.6\times 10^8$~cm in the pre-SN
star. Our fallback calculations find that the outermost radial
distance that this bound shell of material reaches during the
explosion before turning back is $r_{\rm max}\sim 1.8\times
10^{10}$~cm. The free-fall time of this material is on the order of a
few hundred seconds, and its circularization radius is at about
$1.7\times 10^6$~cm.

\begin{figure}[t!]
\begin{minipage}[t]{0.5\textwidth}
\includegraphics[width=8cm]{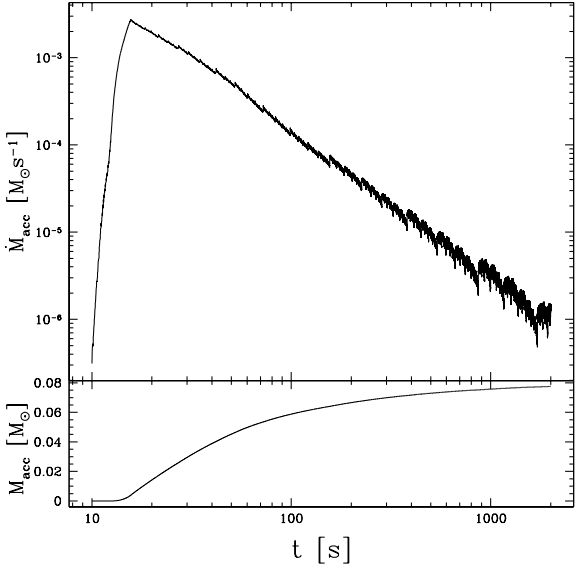}
\end{minipage}
\caption{Accretion rate {\em (top panel)} and total accreted mass 
{\em (bottom panel)} for the Z1M16v20 model with no magnetic torques,
and an explosion energy of $3\times 10^{52}$~erg. For this explosion energy,
only $\sim 0.08 M_\odot$ remains bound and falls back, and the remnant
compact object remains an NS.}
\label{fig:MdotNS}
\end{figure}

The time-dependent fallback accretion rate of the bound $0.08 M_\odot$
shell of material is displayed in Fig.~\ref{fig:MdotNS}. {  The
  temporal power-law decay, $\sim t^{-5/3}$, has been found before in
  similar types of fallback calculations (MacFadyen et al 2001), which
  relied on an alike density profile to ours.}\footnote{  The
    semi-analytical fallback calculations of Kumar et al. (2008) also
    found an initial rate of fallback $\propto t^{-5/3}$, followed by
  a stepeer decay after a few hundeds of seconds, which, in their
  particular progenitor model, corresponded to a drop in the density
  profile. A direct correspondence between the fallback timescale in
  their model and ours (as well as that of MacFadyen et al. 2001)
  cannot be made, since their computation did not include a simulation
  of the explosion and the consequent fallback delay due to the shock
  pushing out the material before it starts to fall back. Also, if the
explosion is strong and most of the material gets unbound
such as in the case of Fig.~\ref{fig:MdotNS}, then the steepening 
in $\dot{M}$ is not seen since the outer parts of the star envelope,
where the density drops, have become unbound.}  Since the
viscous timescale $t_0$ at the circularization radius is very short
(about a millisecond), and hence $t_0\ll t_{ff}$, the disk can be
imagined as evolving through a series of steady-state configurations
at very high accretion rates, varying between $\sim
10^{-3}\,M_\odot$~s$^{-1} - 10^{-6}\,M_\odot$~s$^{-1}$ within a
few hundred seconds. This is an accretion regime known as
hyperaccretion.

Hyperaccretion around an NS, unlike the BH case, has received a rather
limited attention in the literature. Kohri et al (2005) analyzed the
problem with the aim of searching for an extra energy source which may
help supernova explode in numerical simulations. They pointed out how, due
to the very high accretion rates, the accretion flow, once
circularized into a disk, would be a neutrino-dominated accretion flow
(NDAF; Popham et al. 1999; Narayan et al.  2001; Kohri \& Mineshige
2002). Their analysis showed that, under a wide range of conditions,
strong outflows would develop, and carry away a large fraction of
mechanical energy. This energy might then help in producing a successful 
supernova explosion. 

More recently, Zhang \& Dai (2008, 2009) studied the structure of a
steady-state hyperaccretion disk around an NS, motivated by
suggestions that the collapse of a massive star giving rise to a GRB
might be powered by accretion onto an NS instead of a BH.  A
fundamental difference between hyperaccretion onto an NS and onto a BH
derives from the fact that the BH has an event horizon, while the NS
has a hard surface. As a result, if the accreting object is a BH, the
internal energy in the accretion flow can be advected inward into the
event horizon without any energy release. On the other hand, if the
accreting object is an NS, the internal energy must eventually be
expelled from the disk, since the NS surface prevents energy advection
into the star. A hyperaccreting disk around an NS, similarly to the BH
case, is very hot and dense, optically thick to photons, and hence
cooling is dominated by neutrinos.  The presence of the boundary layer
at the NS surface, and the consequent inability of the NS to advect
energy, causes an higher neutrino luminosity than in the case of BH
hyperaccretion. {  Such explosive events could therefore power long
  GRBs. However, a quantitative comparison between the fallback
  accretion rate predicted by our numerical simulations (cfr.
  Fig.\ref{fig:MdotNS}), and the temporal evolution of the luminosity
  observed in long GRBs (e.g. Zhang et al. 2006) cannot be reliably
  made at this stage. As discussed above, the detailed way by which hyperaccretion
  around an NS proceeds is not well known. Furthermore, the efficiency
  by which the fallback accretion rate gets converted into accretion
  rate onto the NS and/or into winds must be changing with time,
  depending on the magnitude of the fallback rate itself, since the
  accretion mode depends on $\dot{M}$, and the accretion rate onto the
NS surface must remain Eddington limited. And this is in addition to the
  uncertainty with which the accretion energy is then converted into
  $\gamma$-rays during the prompt phase. Also note that, if the
 remnant compact object is a rapidly spinning, highly magnetized NS
as suggested as an alternative model for the long GRB engine (Metzger et al. 2011),
then the energy output (and hence the possible signatures in the light curves)
may be dominated by the electromagnetic component rather than the accretion one.}

The above studies highlight the importance of the formation of
hyperaccreting disks around NSs for supernovae and Gamma-Ray Bursts. 
Our investigation shows that such disks can be formed, but under 
special conditions: (i) magnetic breaking must not be effective 
during the progenitor star evolution; (ii) the explosion energy must
be fine-tuned so that a fraction of mass remains bound, but only
a small fraction, not exceeding a few tens of solar mass in order
for the NS to remain stable, or else it would produce an accretion
induced collapse of the NS into a BH, also characterized by an
early hyperaccreting phase (Giacomazzo \& Perna 2012).  

The disks discussed above, independently of their specific structure,
and the presence or not of outflows, are characterized by being highly
transient, with relatively short durations of at most several hundreds of
seconds. Given their formation within a few Schwarzschild's radii,
these disks get rapidly depleted before they can expand to
sufficiently large radii and form the extended, low-accretion, and
long-lived disks ($10^4-10^5$~yr) which have been widely discussed in
the NS literature, as summarized in the introduction.  Therefore, we
can robustly conclude that {\em long-lived fallback disks around isolated
NSs are not expected to be common}. Their formation would require a set
of very fine tuned conditions, such as an explosion which is strong
enough to unbound the innermost layers surrounding the NS, but
anisotropic in such a way that some of the outer, high-$j$ layers were
not affected and could fallback on a much longer timescale at much
larger radii. Mixing between layers with different initial angular
momentum could further occur due to hydrodynamic instabilities
(e.g. Fryxell, Arnett \& Mueller 1991; Chevalier, Blondin \& Emmering
1992).  However, whether these instabilities could possibly lead to
the formation of a small-mass disk at larger radii (with everything
inside emptied to avoid the accretion induced collapse of the NS to a
BH) would have to be investigated through detailed numerical
simulations.

The above considerations have not considered the potentially important
effect of the interaction between the accreting matter and strong
magnetic fields in the newly born NSs. This is a topic which has
received a considerable degree of attention in the literature, both
analytically and semi-analytically (e.g. {  Illarionov} \& Sunyaev 1975;
Ghosh \& Lamb 1979; Lovelace et al. 1995, 1999; Ikhsanov 2002;
Rappaport et al. 2004; Eksi et al. 2005; D'Angelo \& Spruit 2010; Piro
\& Ott 2011), as well as numerically (e.g. Hayashi et al. 1996; Miller
\& Stone 1997; Fendt \& Elstner 2000; Matt et al. 2002; Romanova et
al. 2004, 2009).  A strong magnetic field exerts a pressure on the
accretion flow, which is able to balance the ram pressure of the
accreting material at a radius on the order of the magnetospheric
radius, $R_m = [\mu^2/(2GM)^{1/2}\dot{M}]^{2/7}$, where $\mu$ is the
magnetic stellar moment (e.g. Davidson \& Ostriker 1973). This radius
defines approximately the inner boundary of the accretion disk.  A
magnetized NS is able to accrete only under the condition that, at the
magnetospheric radius, the velocity of the rotating magnetosphere
(equal to the stellar angular velocity $\Omega_0$) is smaller than the local
Keplerian velocity of the disk material, $\Omega_K(R_m)$; if this
condition is not satisfied, a centrifugal barrier will inhibit
accretion ({  Illarionov} \& Sunyaev 1975). 
The above condition can be equally reformulated by stating that
the magnetospheric radius must be smaller than the corotation
radius $R_{co}=(GM/\Omega^2)^{1/3}$, which is the radius at which the
Keplerian frequency of the orbiting matter is equal to the NS spin
frequency. For a hypothetical fallback
disk not to be influenced by the magnetic field of the NS, the magnetospheric
radius must not be larger than the corotation radius, which yields a lower limit
onto the accretion rate, 
\begin{equation}
\dot{M}\, > \dot{M}_{\rm prop, crit}\,= 
6\times 10^{-9}\,\mu_{30}^2\, M_{1.4}^{-5/3}\,P_1^{-7/3}\, M_\odot\, {\rm s}^{-1}\,,
\label{Mdotprop}
\end{equation}
where $\mu_{30}\equiv \mu / 10^{30}$~G~cm$^3$, $P_1\equiv P/1~{\rm s}$, 
$M_{1.4}\equiv M/1.4 M_\odot$. 
Other effects (related to the presence of magnetic fields) which can
hinder the formation of a disk are dipole spin-down
radiation and a neutrino-driven, magnetically-dominated wind
(Thompson et al. 2004; Piro \& Ott 2011). These yield, respectively, further limits 
on the accretion rate, 
\begin{equation}
\dot{M}\, > \dot{M}_{\rm dip, crit}\, = 
1.8\times 10^{-11}\,\mu_{30}^2\, M_{1.4}^{-1/2}\,P_1^{-4}\, R_{12}^{1/2}\, M_\odot\, {\rm s}^{-1}\,,
\label{Mdotdip}
\end{equation}
and
\begin{equation}
\dot{M}\, > \dot{M}_{\nu, \rm crit}\, =
5.4\times 10^{-8}\,\mu_{30}^{4/5}\, M_{1.4}^{-1/2}\,P_1^{-8/5}\,\,\dot{M}_{\nu,-3}^{3/5}\, 
R_{12}^{1/2}\, M_\odot\, {\rm s}^{-1}\,.
\label{Mdotnu}
\end{equation}
In the above equations, $R_{12}\equiv R/12~{\rm km}$, and 
$\dot{M}_{\nu,-3}\equiv \dot{M}_\nu/10^{-3}\,M_\odot\,{\rm s}^{-1}$
is the mass-loss rate caused by the neutrino-driven wind.

In the case of the disks produced under the conditions discussed above
(negligible magnetic torques, strong explosion that leaves only the
innermost few tens of solar masses bound), we find that the very high
accretion rates, $\dot{M}\ga 10^{-6}\,M_\odot/s$, generally dominate over
magnetic effects unless the NS is born with a magnetar-type field, and
with a period on the order of the millisecond. In these cases, the
NS-disk system would be in the propeller phase, and the 
disk may be disrupted through a combination of the transfer of accretion
power and rotational power of the NS to the disk (Eksi et al. 2005).

All together, the presence of strong magnetic fields in the newly born
NS would only make the formation and survival of a hypothetical
fallback disk even more difficult. However, we note that a possible
avenue that might create the conditions for an extended disk at larger
radii could occur through 'recycling' of the matter expelled during
the propeller phase, i.e. if some of the material which has gained
angular momentum through momentum transfer at the magnetospheric
radius during the propeller phase does not get unbound, and falls back
at radii much larger than $R_m$ (Perna, Bozzo \& Stella 2006).  While
this is a possibility, we however note that it requires even more
special conditions to be realized: negligible magnetic torques so that
the innermost layers of material have $j>j_K(R)$, a strong explosion so
that only a few tens of mass fall back; strong NS $B$ field and short
birth period to allow the propeller phase to set in, anisotropic
conditions in the pulsar wind outflow so that the disk is not blown
away, and constraints on the parameters $B$, $P$ and $\dot{M}$ so that a
fraction of the propelled material does not get unbound.

\subsection{Fallback and disk formation around Black Holes}

For those stars which leave behind a BH remnant, the formation of a
fallback disk, while being common in the less physically relevant
$B=0$ case, is also relatively common when magnetic torques play a
significant role in the post main-sequence evolution, unlike the NS remnant case.  The
detailed outcome is however a function of mass, metallicity and
rotational speed of the main sequence star, as well as the explosion energy. 
For a fallback disk to be formed, a fraction of material must remain bound,
and it must possess a specific angular momentum $j_m$ at least as large
as 

\begin{equation} j(R) = \frac{\sqrt{GMR} \left[R^2 -
    2(a/c)\sqrt{GMR/c^2} +(a/c)^2\right]} {R\left[ R^2-3GMR/c^2 +
    2(a/c)\sqrt{GMR/c^2}\right]^{1/2}}\;.
\label{eq:jGR}
\end{equation} 
at the last stable orbit, where $j$ represents the specific
angular momentum of a particle on a corotating orbit of a black hole
of mass $M$ and angular momentum $J=aM$. 

From the results of the grid of simulations that we have run, 
we distinguish two main scenarios for the ensuing observable phenomenology:

\noindent
\subsubsection{Hyperaccreting disks from fast-rotating/chemically homogeneous evolved stars}

In our grid, for metallicities $\la 10\%$ {  of solar}, there is, for each mass
value considered, a critical rotational velocity above which the
evolution proceeds in a chemically homogeneous way (see discussion in
Section~2). In the pre-SN phase, the star is compact and the core is
fast rotating.  For explosion energies $\la 10^{52}$~ergs, we have
found that there is always a non-negligible amount of fallback
mass. The range of rotational velocities for which this fallback
material is able to circularize outside the last stable orbit is an
increasing function of the main sequence star mass, as discussed in
Sec.~2.  The free fall time of the outermost layers is typically on
the order of tens of seconds, giving rise to a phase of rapid and
intense accretion.  These types of post-SN outcomes from massive,
rapidly rotating, low-metallicity stars are believed to be the ones
powering the typical long GRBs, and their properties have been widely
discussed in the literature, both in the context of their formation
within the SN collapse (Woosley 1993, MacFadyen \& Woosley 1999;
MacFadyen, Woosley \& Heger 2001), as well as in the context of the
post-explosion evolution, when a hyperaccreting, neutrino-cooled disk
is believed to be formed (Popham et al 1999; Narayan et al. 2001; Di
Matteo et al. 2002; Janiuk et al 2004, 2007; Lee \& Ramirez-Ruiz 2006;
Surman et al. 2006; Shibata et al. 2007; Chen \& Beloborodov
2007). The duration of these disks, while varying depending on the
specific angular momentum distribution of the progenitor star
(e.g. Janiuk \& Proga 2008), remains within at most several hundreds
of seconds.
 
Given the extensive investigations in the literature on the topic
of hyperaccreting disks around BHs and their relevance to GRBs,
we will not discuss these disks any further here, and refer the reader
to the appropriate literature cited above for details.

\subsubsection{Long-lived disks from slow-rotating/chemically {  inhomogeneous} evolved stars}

As discussed in Section~2, for each couple of $Z,M$ values, there
exists a critical rotational velocity below which the evolution does
not lead to full mixing. These stars 
{  lose} substantially less mass, and hence they 
have a much larger radial extent in the pre-SN
phase. The free fall times of the outermost layers can reach the
$\sim$ year timescale for very low-metallicity, massive stars (see
also Paxton et al. 2013). By inspection of the angular momentum
distribution of the chemically {  inhomogeneous} evolved stars in
Figs.~\ref{fig:preSN1}, \ref{fig:preSN2} and \ref{fig:preSN3}, we note
that, depending on the explosion energy, there is a very large range
of timescales for the bound material to fall back, all the way from
milliseconds to years.  Therefore, transients with durations of
several days such as those observed in the sample of very long GRBs
(Levan et al. 2013) should not be as surprising in the context of
massive star explosions (as long as the jet still manages to penetrate
the envelope). Here, we will not dwell on these intermediate-duration
cases (see Woosley \& Heger 2012 for a more in-depth discussion of
these situations), but we will rather focus on the fate of the outermost
layers of the envelope in those scenarios (weak and/or anisotropic
explosion) in which these outermost layers still remain bound after the
explosion, and eventually fall back.

Whether this material is able to circularize into a disk is, once
again, a function of metallicity, mass, and initial rotational
velocity of the MS star. For the $1\%$ and $10\%$ metallicity cases,
the $13$ and $20~M_\odot$ models display a significant range of
rotational velocities for which the specific angular momentum of the
fallback material is well above the required value to circularize
beyond the last stable orbit.  The velocity range for circularization
is reduced for the higher mass models that we studied ($30$ and
$40~M_\odot$). However, even at solar metallicities, the two lower
mass cases still display a range of rotational velocities for which
the specific angular momentum of the outer envelope exceeds the
circularization value at the last stable orbit.

The fate of these high-$j$ outer layers is heavily dependent on the
explosion energy. Not surprisingly, they are blown away for a wide
range of explosion energies. However, for very weak explosions, the
outer envelope may still remain loosely bound, and hence able to form
a disk upon return.  Our fallback calculations (see Section~3) which,
for the Z1M13v32 and Z10M13v32 models were run for much longer
timescales ($t_{\rm fin}=10^7$~s) and for weaker explosion energies
than displayed in Fig.~\ref{fig:fallback} (down to $E=10^{48}$~erg),
have shown that, for explosion energies $\la$ a few $\times
10^{48}$~erg (Z10M13v32 case) and $\la 10^{49}$~erg (Z1M13v32 case),
the outer layers remain bound.  These scenarios lead to BHs in the
$\sim 10 M_\odot$ range, surrounded by extended fallback disks of
several solar masses.

We remark that the values of the explosion energy that we quote here
and throughout the text represent the isotropic values. However, if the
explosion is anisotropic, as suggested by observations of SN~1987A
(e.g.Hillebrandt \& H\"oflich 1989; McCray 1993) and a number of other
SNe (e.g. Fassia et al. 1998; Spyromilio 1991; Woosley et al. 1994),
and a higher fraction of energy is deposited along the rotation axis,
then, even an apparently rather energetic explosion could result in a
relatively low amount of energy transferred to the equatorial regions,
and hence a larger amount of fallback material would be available to
form a disk.  Alternatively, other avenues to produce fallback of the
outer layers could involve very weak or absent shocks.  Woosley \&
Heger (2012) explored this avenue in the context of blue supergiants
with low mass-loss rates, tidally interacting binaries with either
helium stars or giant stars, and the collapse to a BH of very massive
pair-instability supernovae.

In the following, we will study in more detail the expected properties
of these fallback disks, and their relevance to observations for a
couple of specific cases, in particular models Z1M20v25 and Z10M13v32.
We begin by estimating the initial radial extent of the disk by
computing the circularization radius $R_{circ}$ for the outermost
shells, if they all fall back. In the Z1M13v37 model, the innermost
$\sim 5-6 M_\odot$ of material accretes rapidly onto the BH without
being able to circularize.  Outside of those envelope layers, the
infalling material possess sufficient angular momentum to form a
disk. By solving the equation $j(R_{\rm circ})=j_m$ as discussed
above, we find that the outermost $\sim M_\odot$ of material
circularizes in a ring between $\sim 1.4\times 10^9 - 3.8\times
10^9$~cm for model Z10M13v32, and between $\sim 2.7\times 10^9 -
1.6\times 10^{10}$~cm for model Z1M20v25.  For a BH of $\sim 10-11\,
M_\odot$ in the former case, and of $\sim 17-18\, M_\odot$ in the
latter, these outermost regions correspond to $\sim 1000$
Schwarzschild's radii scale.

In the post explosion phase, the fallback material is very hot, and
$H\sim R$, yielding $t_0\sim 200$~s. The free-fall time is considerably
longer. Even in the case of a negligible shock, i.e. with the outer
layers falling back from their original location within the progenitor
star, $t_{ff}\ga 10^6$~s. Since $t_{ff}\gg t_0$, the initial accretion
rates are on the order of $\dot M_d\sim M_d/t_{ff} \sim 10^{-6}
M_\odot$~s$^{-1}$.  After the envelope has fully fallen back, the
evolution will be driven by viscous evolution. The early times during
which the accretion rate is highly super Eddington are characterized by
a hot, optically thick flow, similar to the transient GRB phase, but
with some important differences, and in particular, that the disk is
no longer neutrino-cooled. Chen \& Beloborodov (2007; see also Popham
et al. 1999) find that, below a certain critical accretion rate
$M_{\rm c}$, neutrinos are no longer dominating cooling.  $M_{\rm
  c}$ is a function of both the BH spin parameter as well as the
viscosity parameter $\alpha$. For $\alpha=0.01$, neutrino cooling is
negligible within $\sim 100 \,R_s$ for any value of the spin parameter
if $M_{\rm c}\la 0.01 \,M_\odot$, and $M_{\rm c}$ increases with
$\alpha$. At the much lower fallback accretion rates expected from the
outer layers of the stars under consideration, neutrino cooling will
hence be unimportant. Under these conditions, accretion becomes
relatively inefficient, and may proceed as an advection-dominated
accretion flow (Narayan \& Yi 1994, 1995). These types of flows,
characterized by a positive Bernoulli constant, lose considerable mass
into a wind, resulting in an accretion rate which decreases with
radius.  The available kinetic luminosity, on the order of
$10^{46-47}$~erg~s$^{-1}$, may however still be sufficient to power a
jet, albeit much weaker than in the standard GRBs, and yield an event
which, at early times, may resemble a weak and very long $\gamma$-ray
transient (Quataert \& Kasen 2012, Woosley \& Heger 2012).
 
The time-dependent, long-term evolution of a hyperaccreting disk
around a BH has been discussed by Cannizzo \& Gehrels (2009; see also
Kumar et al. 2008; Metzger et al. 2008; Cannizzo et al. 2011) in the
framework of the self-similar solutions derived by Pringle (1974,
1991) for the time evolution of disks with a free outer boundary for
the viscosity parameterization (without outflows).  Setting the
precise initial condition for the beginning of the self-similar
evolution (post-fallback phase) is not a trivial issue.  During the
long fallback period, the very short viscous timescale causes previous
rings of fallback material to quickly spread; the inner layers accrete
onto the BH, while the outer edge moves to increasingly outer
radii. Fresh material that falls back may then interact and shock with
the older circularized material rings. In the following, we will focus
on the fate of the outermost, highest $j$ layers of the envelope,
which fall back last and circularize at radii in the range of a few
$\times 10^9$ - a few $\times 10^{10}$~cm.  In the super-Eddington
slim disk model, and for timescales longer than the viscous time
$t_0$, the time evolution of the disk mass $M_d$, of its accretion rate $\dot M_d$, and of its
radius $R_d$ can be approximated by the self-similar solutions
\begin{eqnarray}
M_d (t)  & = & M_d (t_0) \left( \frac{t}{t_0} \right)^{-1/3}\,,\\
\dot M_d (t)  & = & \dot M_d (t_0) \left( \frac{t}{t_0} \right)^{-4/3}\,\\
R_d (t)  & = & R_d (t_0) \left( \frac{t}{t_0} \right)^{2/3}\,,
\label{eq:evol1}
\end{eqnarray}
under the assumption of negligible mass outflows (and hence
radius-independent mass accretion rate; Kumar et al. 2008; Metzger et
al. 2008; Cannizzo et al. 2011).  At this rate of accretion, the flow
becomes sub-Eddington after about 10~years (scaling in units of a BH
mass of 15~$M_\odot$ and a ring initially at $10^{10}$~cm). At that
time, the left-over disk mass (for an initial ring of $\sim
1~M_\odot$) is about a percent of solar mass, and the outer radius of
the disk has expanded to about $10^{14}$~cm.  From that point on, the
disk can become easily cooled by photons, and the structure is more
likely to resemble an optically thick, geometrically thin disk. With
$H/R\sim 0.1$, the viscous timescale becomes now substantially longer,
$\sim$ a few $\times 10^{8}$~s, and the accretion rate declines at a
slightly more gentle rate. During this phase of evolution, Cannizzo et
al (1990) found that the disk mass, the accretion rate and the outer
radius can be approximated by the self-similar solutions
\begin{eqnarray}
M_d (t)  & = & M_d (t_0) \left( \frac{t}{t_0} \right)^{-3/16}\,,\\
\dot M_d (t)  & = & \dot M_d (t_0) \left( \frac{t}{t_0} \right)^{-19/16}\,,\\
R_d (t)  & = & R_d (t_0) \left( \frac{t}{t_0} \right)^{3/8}\,,
\label{eq:evol2} 
\end{eqnarray}
for $t\ga t_0$.  Note that, in both the above self-similar solutions, the
equation for the disk radius results from the assumptions that most of
the disk mass resides in its outermost regions and that the total
angular momentum of the disk is conserved. Both these conditions are
satisfied in the disk after the fallback phase has ended.

According to the relations above, the disk has the potential to be
bright for a very long time. For example, at an age of $\sim 10^5$~yr,
such a disk, of mass $\sim 10^{-3}~M_\odot~{\rm s}^{-1}$, would have an
accretion rate of about $\sim 10^{14}~{\rm g}~{\rm s}^{-1}$.  The
expected luminosities of these disks for ages $\sim 10^4-10^5$~yr end
up to be in the range of what was discussed for fallback disks around
NSs as models to explain the luminosities of the Anomalous X-ray
Pulsars (Perna, Hernquist \& Narayan 2000). We should also note that the
potential for luminous fallback disks left over around BHs after the
NS explosion has been discussed by Liu (2003) as a model for
Ultraluminous X-Ray Sources.

After the last phase of a geometrically thin, optically thick disk has
set in, the following evolution of the gaseous material will depend
heavily on the characteristics of the opacity, and in turn on the
metallic content of the material.  A detailed discussion of fallback
disks from SN explosions (albeit aimed at disks formed around NSs) was
presented by Menou et al. (2001b).  Their investigation revealed that
the powerlaw evolution continues until a point at which the
temperature becomes low enough that the thermal-ionization instability
may set in (e.g. Haumery et al. 1998; Menou et al. 2001b).  From that
point on, the magneto-rotational instability, which provides an
important source of viscosity (Balbus \& Hawley 1991) may no longer be
able to operate, and then the outcome would be that of a 'dead' disk,
unable to accrete viscously, and simply cooling. For a solar
composition disk, the local stability criterion is (e.g. Hameury et
al. 1998)
\begin{equation}
\dot M_d(R) > \dot M_{\rm crit}(R) \simeq 9.5\times 10^{15}\,
m_{10}^{-0.9}\,R_{10}^{2.68}\, {\rm g}~{\rm s}^{-1}\,,
\label{eq:Mcrit}
\end{equation}
where $R_{10}$ is the radius of interest in units of $10^{10}$~cm.
For Helium and metal-dominated disks, one may expect slight deviations
from the scaling above (e.g. Menou et al. 2001b).  In the case of the
disks under consideration here, and strictly adopting
Eq.~(\ref{eq:Mcrit}), the outermost parts would have become passive up
to a radius of $\sim 2\times 10^{11}$~cm at the time of the transition
between the super-Eddington phase described by Eq.~(\ref{eq:evol1}) and
the sub-Eddington one described by Eq.~(\ref{eq:evol2}). However, we
note that Eq.(\ref{eq:Mcrit}) has been derived under the condition of
a geometrically thin disk, and hence it should not be applied to the
optically thick, hyper Eddington conditions of the early
times\footnote{The stability properties of hyperaccreting disks have
  been studied by a number of authors (e.g. Di Matteo et al. 2002;
  Chen \& Beloborodov 2007); due to the very high temperatures, these
  disks have been found to be stable with respect to the thermal and
  viscous stability, but possibly gravitationally unstable in the
  outer regions (see also Perna, Armitage \& Zhang 2006). Dead zones
  due MRI suppression have been identified as a result of the neutrino
  viscosity in the innermost parts of the disk (Masada et
  al. 2007). However, these are not of a concern to us, since at
  accretion rates $\sim 10^{-7}-10^{-6} M_\odot$~s$^{-1}$ the flow is
  optically thin to neutrinos.}.  Furthermore, even during the early
phases of the sub-Eddington phase, when the accretion rate is still
relatively high, irradiation of the outer parts of the disk by the
X-rays produced in the inner regions of the disk itself (disk
self-irradiation) can affect both the time evolution of the accretion
rate ($\dot M_d\propto t^{-4/3}$ for irradiation-dominated disks;
Cannizzo \& Gehrels 2009), as well as the critical mass accretion rate
below which the disk becomes thermally unstable (Dubus et al. 1999;
Menou et al. 2001b; see also Currie \& Hansen 2007 for fallback disk
evolution with layered accretion).

In the disks under consideration here, the outer parts have had enough
time to expand to relatively large radii ($10^{13}-10^{14}$~cm) during
the hyper Eddington phase, when the viscous timescale is very short.
Therefore, albeit keeping in mind that the scalings of Eq.(\ref{eq:evol1}) 
are approximate, and that there are many uncertainties in the 
early hyper Eddington phases (including the likely presence of outflows which
are not captured by Eq.(\ref{eq:evol1})), we entertain the possibility
that, early on, mass has been transported beyond the radius at which
the MRI stops operating when the disk cools at later times.

Among the interesting consequences of a long-lived and extended
fallback disk around a BH, is the intriguing possibility of forming
planets, which, after the disk has dispersed, would remain in orbit
around the BH.  Such a scenario was originally proposed by Lin et
al. (1991) for the origin of the planets around the radio pulsar PSR
1257+12 (Wolszczan \& Frail 1992; Wolszczan 1994), and further
investigated by Currie \& Hansen (2007).  Conditions for planet
formation, through phases of grain growth and planetesimal formation,
have {  since} been shown to arise in the outer regions of these fallback
disks. However, for planets to be able to survive, the disk must
extend beyond the tidal disruption radius for a self-gravitating body
of density $\rho$ (e.g. Aggarwal \& Oberbeck 1974) orbiting an object
of mass $M$
\begin{equation}
R_{\rm tid} \approx \left( \frac{M}{\rho}\right)^{1/3}\,.
\label{Rtid}
\end{equation}
For a BH of $\sim 15~M_\odot$, a self-gravitating body of density 
$\rho \approx 1$~g~cm$^{-3}$ has a tidal radius of $\sim 2.5\times 10^{11}$~cm,
and hence the outer layers of mass which have spread out to $\sim$~AU distances
would allow for the formation and survival of planets. 

\section{Summary}

Fallback disks around newly born compact objects have been invoked to
explain a large variety of phenomena. Hyperaccreting disks from the
collapse of low-metallicity, rapidly-rotating stars are believed to
power long gamma-ray bursts, and have been amply discussed in the
literature. In this paper, using the state-of-the-art open source code
MESA, we explored the fate of fallback material for 60 stellar models with initial mass in the range
$13-40~M_\odot$, metallicities between 0.01 and
1~$Z_\odot$ and initial surface rotational velocities between 0.25 and 0.75 of the
critical value.  Our main focus has been the fate of the high-$j$
fallback material able to circularize and form a disk.  By considering
a range of explosion energies, we have explored fallback disk formation
around both NS and BH remnants. Our 'standard' model included transfer
of angular momentum among various layers via magnetic coupling;
however, we also explored the evolutionary scenario under the
assumption of negligible magnetic torques.

Our main findings are summarized below:

\begin{itemize}

\item
For strong explosions which leave behind NS remnants, we find that, if
magnetic coupling plays an important role during the stellar evolution, 
the specific angular momentum of the shells immediately outside of the iron
core is never large enough for the material to be able to circularize
in an orbit outside of the NS radius. Hence fallback disks around
newly born NSs are not a common outcome of the SN explosion for strong
magnetic coupling.

\item 
If the evolution lacks magnetic coupling among the various layers, a
much higher $j$ in the pre-SN star is obtained. In this case, the
fallback material does have enough angular momentum to circularize
outside the NS radius, even for solar metallicity stars.  However, the
distribution of the specific angular momentum would require rather
fine-tuned conditions in order to produce an extended, long-lived
fallback disk. For a very narrow range of explosion energies, a few
tens of solar mass of material can fallback and circularize, but just
outside the NS surface, hence resulting in a hyperaccreting, highly
transient and short-lived disk. Circularization of the outer layers of
the envelope at some larger radii would require very special
conditions in the explosion geometry: the innermost layers would need
to be blown away (or else the NS would quickly turn into a BH), while
the outermost layers would need to receive very little of the
explosion energy.  These special conditions, coupled with the fact
that models without magnetic torques are disfavored by observations
of the periods of young pulsars, have led us to conclude that {\em
  extended, long-lived fallback disks around isolated NSs are not
  expected to be a common outcome of supernova explosions.}

\item 
For explosion energies leading to BH remnants, we find that there is
always a region in the $M,Z,v,E$ parameter space for which fallback
disks are a possible outcome, independently of whether magnetic
torques play an important role in the evolution of the star. However,
the size of these regions in the parameter space depends on the
strength of the magnetic coupling.  The explosion material from stars
evolved with negligible magnetic torques is able to circularize even
for solar-metallicity models with typical initial surface velocities
$\vsurfi \sim 0.2\,\vcrit$. In the models with strong magnetic
coupling, on the other hand, fallback disks are common predominantly
at lower metallicities (consistent with what has been found by a
number of previous investigations).

\item
The lower metallicity, faster rotating stars, which lose the outer
envelope during their evolution, possess compact fallback disks
(circularization occurs within a few tens of Schwarzschild' radii),
which result in highly super Eddington accretion rates. The typical
fallback time is on the order of a few tens of seconds. The properties
of these disks have been widely investigated in the literature as the
engines powering long duration Gamma-Ray Bursts, and we have briefly
reviewed them here.

\item
For very weak explosions (or largely anisotropic explosions, with
little energy released in the equatorial regions), even the outermost
layers of the star can remain bound. In the slower rotating models,
for masses $\la 20 \,M_\odot$ and metallicities $\la \,0.1 Z_\odot$, these
outer layers possess a large specific angular momentum.  When they
fall back, a relatively large mass, $\sim M_\odot$, circularizes at a
large radius, $\ga 10^3 R_s$.  This outcome is common independently of
the strength of magnetic torques, as long as the explosion is
sufficiently weak or sufficiently asymmetric.

These massive disks around BHs are expected to be long-lived, although
the specifics are largely dependent on the properties $Z,M,v,B$ of the
main sequence star.  For a typical massive star of $M=13\,M_\odot$,
$\vsurfi \sim 0.3 \,\vcrit$ and metallicity in the range $Z\sim
1-10\,\%Z_\odot$, we find that the outermost $\sim M_\odot$ of material
falls back on a timescale of several months to $\sim$~year, and
circularizes at radii on the order of a few $\times 10^9-10^{10}$~cm
(larger values correlate to lower metallicities of the main sequence
star).  The initial accretion rate is $\sim
10^{-7}-10^{-6}~M_\odot$~s$^{-1}$. During the early hyper Eddington
phase, when the viscous timescale is very short, these disks can
extend to radii on the order of the AU, where tidal forces would not
be strong enough to disrupt large gaseous bodies.  These disks may
harbor the conditions to form planets, hence leading to the possible
existence of planets orbiting BHs.

\end{itemize} 

\acknowledgments

This work was partially supported by NSF grant No. AST 1009396 (RP),
NSF grant AST-1009863 (AM) and NSF grant PHY11-25915.

\end{document}